\documentclass[conference]{IEEEtran}
\usepackage{cite}
\usepackage{amsmath,amssymb,amsfonts}
\usepackage{algorithmic}
\usepackage{graphicx}
\usepackage{textcomp}
\usepackage{xcolor}
\usepackage[caption=false,font=footnotesize]{subfig}
\usepackage{hyperref}
\usepackage{balance}
\usepackage{orcidlink}
\usepackage{listings}
\usepackage{booktabs}
\usepackage{makecell}
\usepackage{multirow}
\usepackage{ulem}

\IEEEoverridecommandlockouts
\def\BibTeX{{\rm B\kern-.05em{\sc i\kern-.025em b}\kern-.08em
    T\kern-.1667em\lower.7ex\hbox{E}\kern-.125emX}}

\begin{document}

\title{BackCache: Mitigating Contention-Based Cache Timing Attacks by Hiding Cache Line Evictions
\thanks{This paper was rejected by \textit{IEEE Transactions on Computers} in 2022,
and \textit{Journal of Systems Architecture} in 2023.
Compared to the previous version,
we have updated the CPU model from \textit{TimingSimpleCPU} to \textit{DerivO3CPU}
to align with more realistic scenarios in modern computer systems,
resulting in a complete re-evaluation of performance metrics.}
}

\author{\IEEEauthorblockN{Quancheng~Wang\IEEEauthorrefmark{2},
                          Xige~Zhang\IEEEauthorrefmark{3},
                          Han~Wang\IEEEauthorrefmark{2},
                          Yuzhe~Gu\IEEEauthorrefmark{2},
                          Ming~Tang\IEEEauthorrefmark{1}\IEEEauthorrefmark{2}}
\IEEEauthorblockA{
    \IEEEauthorrefmark{2} Key Laboratory of Aerospace Information Security and Trusted Computing,
                          Ministry of Education,\\
                          School of Cyber Science and Engineering, Wuhan University, Wuhan, 430072, China\\
    \IEEEauthorrefmark{3} Beijing Smartchip Microelectronics Technology Co., Ltd., Beijing, 100192, China
}
\IEEEauthorblockA{wangquancheng@whu.edu.cn, zhangxige@sgchip.sgcc.com.cn, \{han.wang, yuzhegu\}@whu.edu.cn, m.tang@126.com}
}

\maketitle

\begin{abstract}
Caches are used to reduce the speed gap between the CPU and memory
to improve the performance and efficiency of modern processors.
However, attackers can use contention-based cache timing attacks to steal sensitive information
from victim processes through carefully designed cache eviction sets.
L1 data cache attacks are widely exploited and pose a significant privacy and confidentiality threat.
Existing hardware-based countermeasures mainly focus on cache partitioning, randomization,
and cache line flushing, which unfortunately either incur high overhead or
can be circumvented by sophisticated attacks.
In this paper, we propose a novel hardware-software co-design called BackCache
with the idea of always achieving cache hits instead of cache misses to
mitigate contention-based cache timing attacks on the L1 data cache.
BackCache places the evicted cache lines from the L1 data cache
into a fully associative backup cache to hide the evictions.
To improve the security of BackCache,
we introduce a randomly used replacement policy (RURP) and a dynamic backup cache resizing mechanism.
Our evaluation on the gem5 simulator shows that BackCache 
can degrade the performance by 2.61\%, 2.66\%, and 3.36\% For
OS kernel, single-thread, and multi-thread benchmarks.
\end{abstract}


\section{Introduction}
Caches are used to reduce the speed gap between CPU and memory
to improve the performance and efficiency of modern processors.
Although many security mechanisms are implemented in hardware and software
to protect the confidentiality of the computer system's security-critical information,
there still exists vulnerabilities known as cache side-channel attacks~\cite{lyu2018survey,shen2021micro,su2021survey},
which allows attackers to steal sensitive information
(e.g. user activities~\cite{wang2019unveiling,shusterman2019robust,cook2022there,zaheri2022targeted},
encryption keys~\cite{yarom2014flush+,mantel2017systematic,mushtaq2020winter}),
break security mechanisms (e.g. KASLR~\cite{canella2020kaslr}, Intel SGX~\cite{chen2019sgxpectre}),
and launch various attacks (e.g. Meltdown~\cite{Lipp2018meltdown}, Spectre~\cite{Kocher2019spectre}).
Meanwhile, these attacks are easy to launch and effective on all systems using caches,
either small embedded systems or large cloud servers.

The basic rationale of cache timing attacks is to exploit the timing difference in cache accesses
between cache hits and cache misses to infer the confidential information of victim processes.
We focus on contention-based cache timing attacks~\cite{osvik2006cache,liu2015last},
where the attacker and the victim share the same cache (e.g. L1 data cache, LLC cache),
but they do not share the same memory region during the execution of the attacker and victim processes.
In such types of attacks, the attacker first designs a cache eviction set
and fills the cache to evict the cache lines from one or several specific cache sets,
which results in cache misses when the victim accesses the cache sets.
Then the victim executes its instructions and may access the shared cache lines.
Afterward, the attacker re-accesses the cache lines with the same eviction set used before
and measures the access time of previously evicted cache sets
to learn the access pattern of the victim process and infer the secret information.

Existing mitigation techniques towards contention-based cache timing attacks mainly focus on
cache partitioning~\cite{wang2007new,domnitser2012non,wang2016secdcp,gaudin2023work},
randomization~\cite{qureshi2018ceaser,werner2019scattercache},
and cache line flushing~\cite{zhang2013duppel,oleksenko2018varys,ge2019time,li2022fase}.
Cache partitioning is a hardware-based technique that prevents sharing of cache sets between different processes,
thus providing spatial isolation between the attacker and the victim.
However, this method results in increased latency because of the reduced effective cache size per process,
and it does not apply to the L1 data cache, which is typically limited in size.
Besides, randomization is another hardware-based technique that
randomizes the mapping between the cache sets and memory addresses,
making the contention for cache sets non-deterministic.
However, not only does implementing randomization bring expensive hardware and software modifications,
but also researchers have found that this solution can be bypassed by a sophisticated attack~\cite{song2021randomized}.
Moreover, cache line flushing can be implemented in hardware or software to flush the cache lines
during preemption, process switches, and system calls,
which ensures temporal isolation between the attacker and the victim.
Nevertheless, this approach may incur greater performance overhead,
as the size of the L1 data cache is getting larger (up to 128KB in the latest Apple M1 chip).

On the one hand, these solutions either introduce significant performance overhead or have security concerns,
which may not be suitable for the L1 data cache.
Hence, we need to find a security and efficient solution to retain the benefits of the shared cache,
i.e., allowing each process to access the entire cache and improving the cache hit ratio.
On the other hand, the root cause of contention-based cache timing attacks
is the sharing of cache sets between the attacker and victim processes,
and the deterministic mapping from memory addresses to cache sets,
resulting in deterministic cache line evictions
when the attacker and victim contend for the same cache set.
Therefore, we aim to break this deterministic cache line evictions,
i.e., making the attacker unable to learn whether a cache line is evicted or not.

In this paper, we propose a novel hardware-software co-design called BackCache
to prevent contention-based cache timing side-channel attacks on the L1 data cache.
The main idea of BackCache is to achieve always cache hits instead of cache misses
and hide the cache line evictions from the L1 data cache.
BackCache places the evicted cache lines from the L1 data cache into a fully associative backup cache.
Then the attacker encounters all cache hits when re-probing the eviction set,
without being able to learn the access pattern of the victim process.
Meanwhile, we leverage a custom cache microarchitecture
(i.e., extend a used bit in the cache tag)
and a new \texttt{BUCLR} instruction to support a new replacement policy
called random used replacement policy (RURP) for the backup cache,
which is used to prevent cache lines not reprobed by the attacker from being evicted.
Besides, to prevent attackers from mounting the contention-based attack
on single cache sets with the known size of the backup cache and cache ways,
we design a mechanism to dynamically resize the backup cache with an enabled bit in the cache tag,
which makes it difficult for attackers to infer the secret cache set.
To support this dynamic resizing mechanism, three hardware registers called memory access count register,
minimum backup cache size register and maximum backup cache size register are introduced.

The security and performance evaluation of BackCache is conducted on the gem5 simulator.
The evaluation result demonstrates that our defenses can prevent
contention-based cache timing attacks under several attack scenarios.
Moreover, the estimation using CACTI 6.5 shows that
BackCache introduces 1.77x energy consumption for the L1 data cache
for the whole cache hierarchy compared to the baseline system.
In addition, the overhead measured by LMBench 3.0
increases by about 2.61\% during the context switching.
Lastly, performance evaluation using Mibench
and SPEC 2017 shows an average overhead of 3.80\% and 2.66\% for single-thread user programs.
And multi-thread user program evaluation using PARSEC 3.0
shows an average overhead of 3.36\%.

In summary, we make the following contributions:
\begin{itemize}
\item We propose a new hardware-software co-design called BackCache
to mitigate contention-based cache timing attacks on the L1 data cache
by placing the evicted cache lines into a fully associative backup cache,
along with a random used replacement policy (RURP) and
a dynamic backup cache resizing mechanism.
\item We theoretically analyze the security of fully associative backup cache,
RURP replacement policy, and dynamic backup cache resizing mechanism.
\item We extend the gem5 simulator to implement the BackCache design,
and modify the gcc compiler and Linux kernel to support the instruction extensions.
Our experimental results validate that BackCache can counteract contention-based
L1 cache side-channel attacks while only introducing minor performance overheads.
\end{itemize}

\section{Background and Related Work}
In this section, we provide pertinent background information
on cache timing side-channel attacks and countermeasures.

\subsection{Cache Timing Side-Channel Attacks}
Cache timing attacks exploit key-dependent memory accesses associated with caches
(such as the L1 data cache~\cite{zhang2013duppel,oleksenko2018varys,ge2019time,li2022fase})
to learn secret information.
This interaction is particularly common in software implementations of cryptographic algorithms,
such as the substitution box (S-box) in DES and AES, and the multiplication table in RSA.
In a contention-based cache timing side-channel attack
(e.g. Prime+Probe~\cite{osvik2006cache}, Evict+Time~\cite{osvik2006cache}),
there is no shared memory region between the attacker and the victim, 
which results in one's cache line being evicted by the other.
The attacker can learn the access pattern of the victim by observing the cache hits/misses
when re-probing the eviction set.

\subsection{Existing Mitigation Techniques}
First, cache partitioning isolates cache lines between processes or
security domains to prevent cache timing attacks,
such as RPCache~\cite{wang2007new}, NoMo cache~\cite{domnitser2012non}
and SecDCP~\cite{wang2016secdcp}.
Although cache partitioning defends against contention-based attacks,
it reduces the effective cache per process execution
and does not apply to the L1 data cache due to the limited cache size.
Then, another class of defense explores randomized memory-to-cache mapping in the cache hardware
~\cite{qureshi2018ceaser,werner2019scattercache}.
On the one hand, unfortunately, these randomization-based caches usually result in significant hardware and software modifications.
For example, CEASER~\cite{qureshi2018ceaser} adds a mapping table in hardware
and groups tasks with operating system modification,
while ScatterCache~\cite{werner2019scattercache} adds an index table in hardware,
extends the page table and adds additional communication between kernel and user space by OS support.
On the other hand, randomization-based caches prove to be vulnerable to sophisticated attacks~\cite{song2021randomized}.
In addition, cache line flushing is a recently popular defense against cache timing attacks on private L1/L2 caches.
Simple methods~\cite{zhang2013duppel,oleksenko2018varys,ge2019time} flush the cache when
the processor is about to run another user thread or kernel thread,
while compound flush methods~\cite{bourgeat2019mi6,li2020simf,wistoff2020prevention}
implemented on RISC-V processors flush several microarchitecture states (including L1 caches, TLB, and BPU).
FaSe~\cite{li2022fase} proposes a fast selective flush method to mitigate cache timing attacks on the L1 data cache
and provides similar security guarantees as naive flushing with a lower overhead.

\section{Threat Model}
The threat model considered has a separate attacker and victim process,
and both of them run on the same processor core and share the same L1 data cache,
which has been already investigated in prior work~\cite{zhang2013duppel,oleksenko2018varys,ge2019time,li2022fase}.
We also assume that the attacker and the victim
use different time slices on the same physical core,
a setup consistent with several existing studies on L1 data cache attacks
~\cite{zhang2013duppel,li2022fase}.
This assumption is reasonable, 
as Zhang et al.~\cite{zhang2013duppel} note that SMT may be disabled
in production systems due to security and accounting concerns.
Additionally, many modern processors,
such as the ARM Cortex-A72, A75, and A76 commonly found in mobile devices,
do not support SMT.
In addition, the attacker cannot interfere with the victim's execution
but can partially control the victim to execute the target function
(e.g., AES encryption), and then observe the cache state after context switching occurs.

\begin{figure}[h!]
\centering
\includegraphics[width=2.5in]{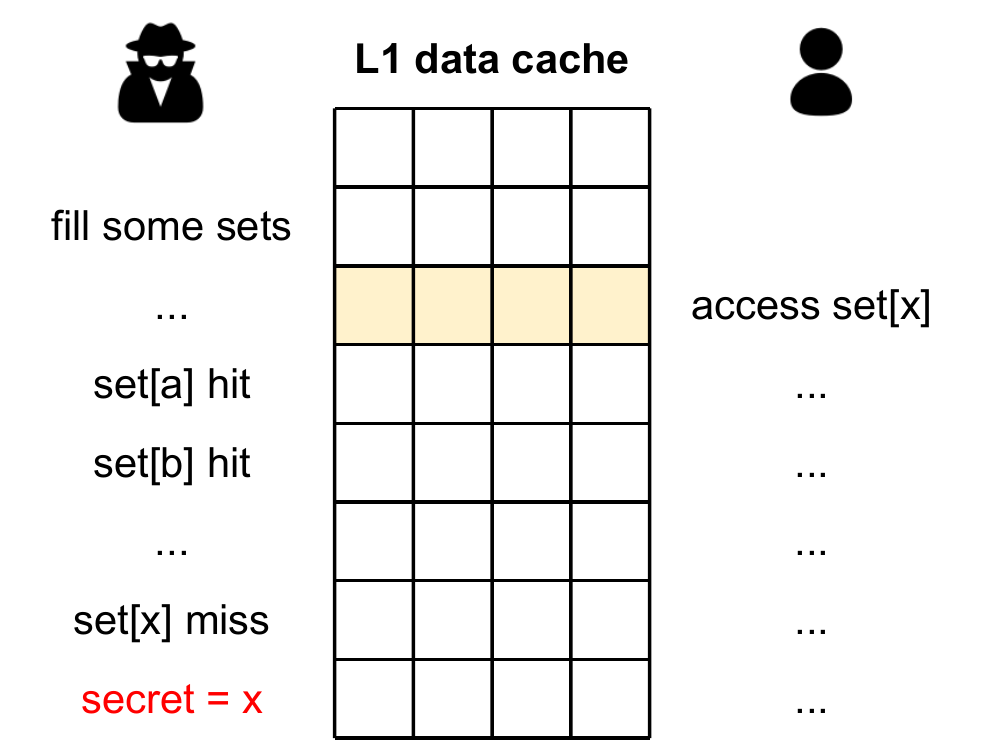}
\caption{Contention-based cache timing attack on L1 data cache.}
\label{fig-threat-model}
\end{figure}

Figure~\ref{fig-threat-model} shows an instance of the contention-based attack
when accessing the L1 data cache shared between attacker and victim.
And following is the sequence of this attack:
\begin{itemize}
\item The attacker and the victim share the same L1 data cache,
and the attacker can access any of the cache sets and contend with the victim.
Then the victim's access to the cache set depends on or is indexed by the secret data.
\item The attacker fills a cache set or some cache sets from the L1 data cache
with a carefully designed eviction set, and then waits for the victim to access the shared data cache.
\item After the victim's execution,
the attacker then accesses the same eviction set and determines
which cache sets have also been accessed by the victim based on the cache hits/misses.
\end{itemize}

Based on our threat model, the attacker can launch two different types of attacks.
First, the victim may only access one cache set during execution
such as a covert channel attack exploited in Spectre attacks~\cite{schwarz2019netspectre}
that encodes the access as “1” and the non-access as “0”,
and the attacker can infer whether the victim performs a memory access according to the access time.
Second, the victim may access multiple cache sets during execution,
such as the Prime+Probe attack on AES encryption,
and the attacker can infer which cache set the victim accesses based on the access time.

\section{Microarchitecture Design}
In this section, we introduce the microarchitecture changes required to implement BackCache.

\subsection{BackCache Architecture}

\begin{figure}[h!]
\centering
\subfloat[Memory hierarchy]{\includegraphics[width=2.5in]{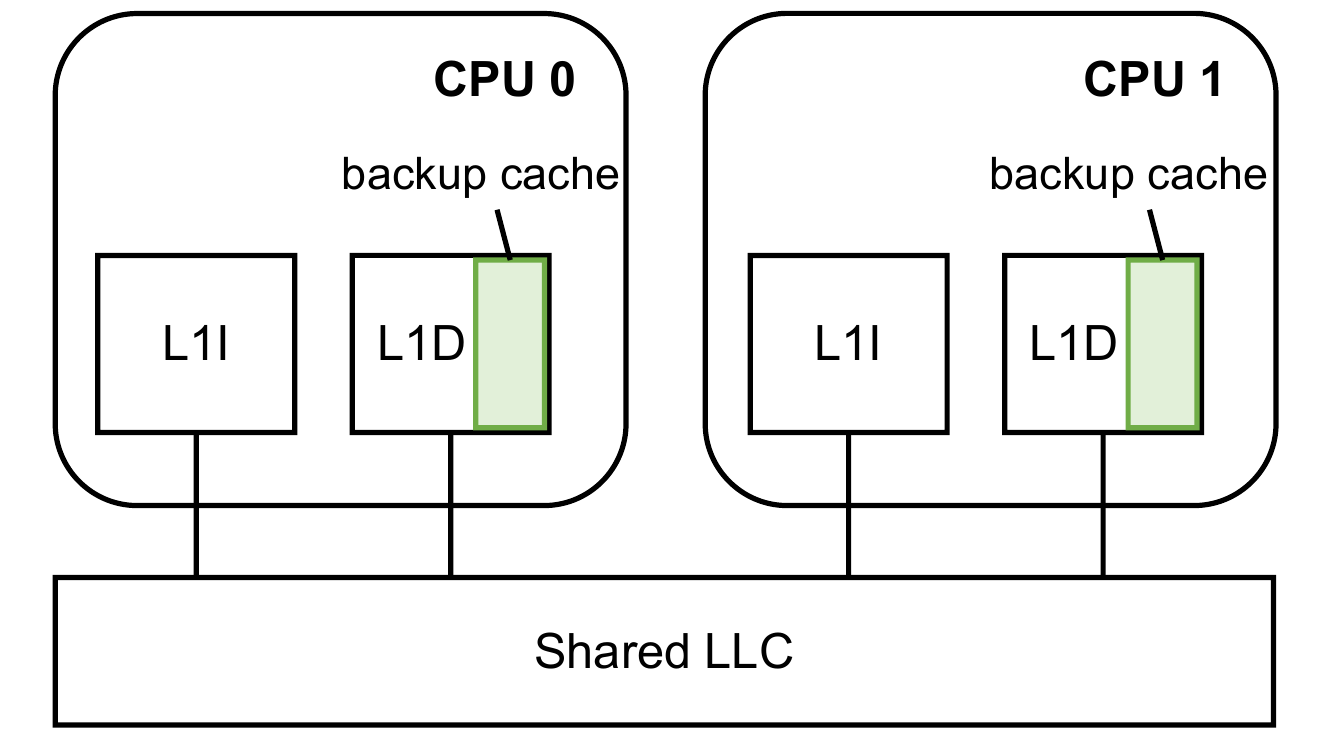}}\\
\subfloat[Backup cache tags (48-bit address space, PIPT)]{\includegraphics[width=2.5in]{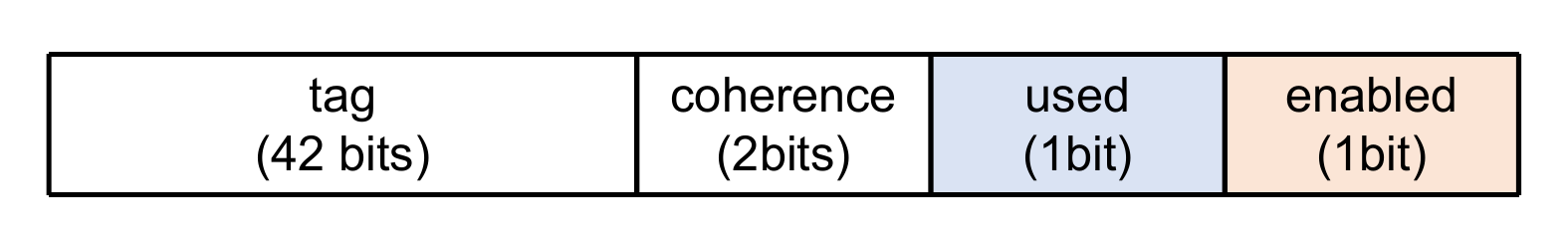}}\\
\subfloat[BackCache overview]{\includegraphics[width=2.5in]{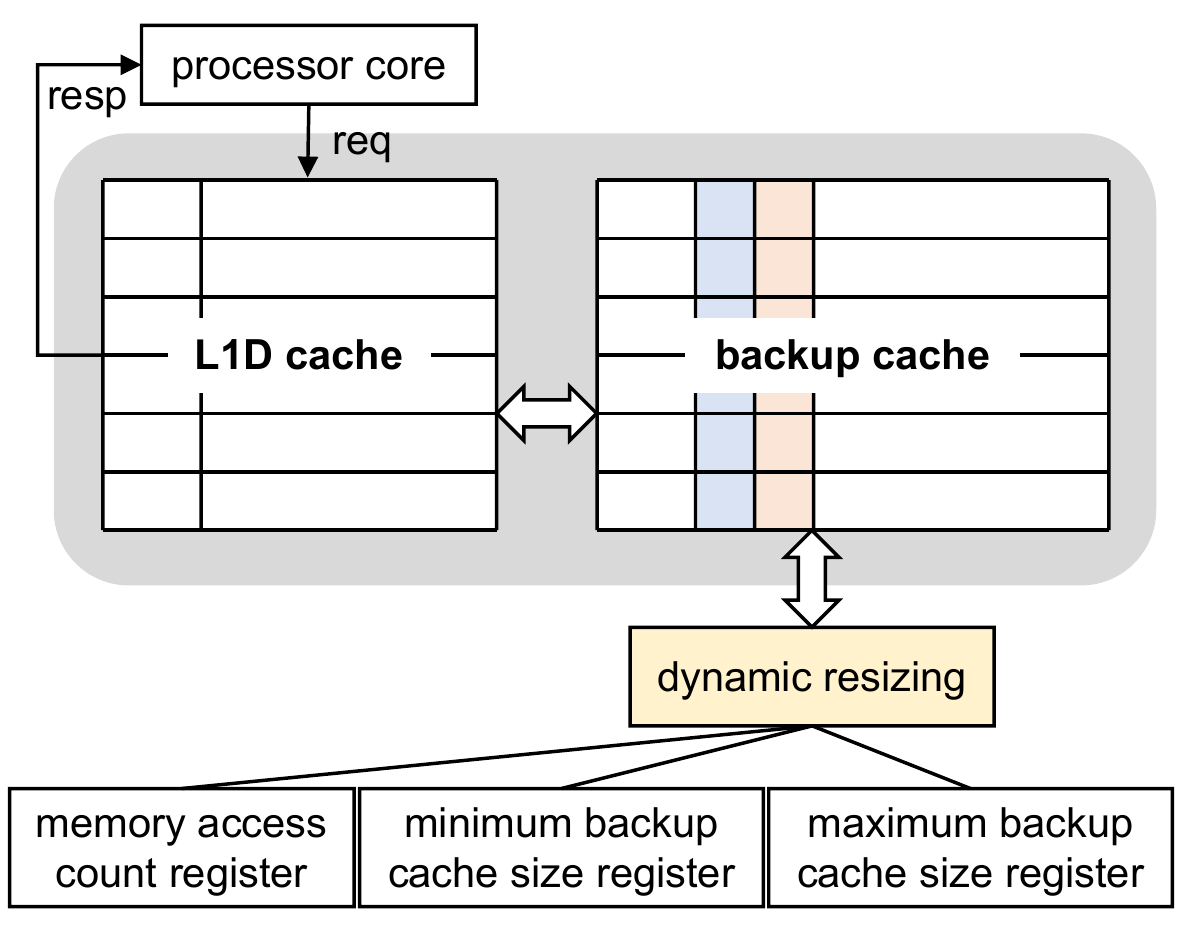}}
\caption{A high-level overview of the BackCache and hardware modifications.}
\label{fig-overview}
\end{figure}

Modern computers typically employ an architecture with
multiple processor cores and hierarchical caches.
Given that existing partitioning and randomization strategies
have effectively ensured the security properties of the LLC,
there is still a deficiency in security assurance for the L1 data cache.
Therefore, filling this security gap and addressing the side-channel vulnerabilities
on the L1 data cache is crucial for ensuring the security of whole multi-core systems.

To achieve the goal of mitigating contention-based cache timing attacks
on L1 data cache by always having cache hits instead of cache misses,
we propose a new cache structure called BackCache,
and Figure~\ref{fig-overview} provides a high-level overview of
its architecture and hardware modifications.
When the processor forwards a memory access request to the L1 data cache, 
the L1 data cache checks both the L1 data cache and the backup cache,
and eventually returns the response to the processor.
The hardware support for BackCache is listed as follows:
\begin{itemize}
\item A fully associative backup cache is added at the same level as the L1 data cache.
\item A used bit and an enabled bit are extended to each cache line in the backup cache.
\item Three hardware registers are added to the processor core:
memory access count register, minimum backup cache size register and maximum backup cache size register.
\item A new replacement policy called random used replacement policy (RURP) is proposed for the backup cache.
\item A dynamic backup cache resizing mechanism is introduced to adjust the size of the backup cache.
\end{itemize}

\textbf{Fully Associative Backup Cache.}
First, we implement the backup cache at the same level as the L1 data cache,
which also has the same access latency as the L1 data cache.
When the processor performs a memory access,
both the L1 data cache and the backup cache can be queried at the same time,
and a hit on either one will result in an L1 cache hit. 
This means that even if the cache line does not exist in the L1 data cache,
the processor still observes an L1 cache hit as long as the data exists in the backup cache.

Then, the backup cache is designed as a fully associative structure.
The key idea of contention-based cache timing attacks is that
the attacker can determine the victim's secret data by observing cache hits/misses on different cache sets.
Therefore, the attacker cannot distinguish the mapping between the L1 data cache and the backup cache,
as long as there is only one cache set in the backup cache,
i.e., the cache line in any cache set of the L1 data cache can be mapped to any cache line in the backup cache.

\textbf{Used Bit and Enabled Bit.}
The used bit extended in the tag array of the backup cache is used to indicate
whether the processor accesses the cache line after being placed in the backup cache,
which is intended to support the random used replacement policy (RURP).
The used bit is set to 0 when the cache line is brought into the backup cache for the first time,
and then set to 1 when it is accessed again, regardless of the presence of the data in the L1 data cache.

The enabled bit is also extended in the tag array of the backup cache,
which is designed to indicate whether this cache line is active in the backup cache
and used for dynamic backup cache resizing mechanism.
If the enabled bit is set to 0, this cache line is disabled in the backup cache,
meaning that no data can be stored to or loaded from this cache line at the current time.
The cache line will not be available again until the enabled bit is set to 1.

\textbf{Hardware Registers.}
To support the dynamic backup cache resizing mechanism,
we also introduce three hardware registers in the processor core: the memory access count register,
the minimum backup cache size register and the maximum backup cache size register.
First, the memory access count register is used to count the number of memory accesses
(no matter cache hits or cache misses) since the last reset of the memory access count.
Second, the minimum backup cache size register and the maximum backup cache size register
are used to indicate the minimum and maximum size of the backup cache,
which are used to control the lower and upper limits of the backup cache size.

\textbf{Random Used Replacement Policy (RURP).}
A distinctive feature of contention-based cache side-channel attacks is that
the attacker re-accesses the eviction set that has been accessed before.
Besides, the used bit is to indicate whether the cache line has been accessed again,
and the used bit of re-accessed cache line by the attacker is more likely to be set to 1.
Therefore, we propose a random used replacement policy (RURP) for the backup cache,
which prioritizes the replacement of cache lines with used bits set to 1
to prevent cache lines that have not been reprobed by the attacker from being replaced in the backup cache,
and thus counter the contention-based cache timing attacks.

\textbf{Dynamic Backup Cache Resizing Mechanism.}
It remains possible for the attacker to determine the victim's secret data
by observing the cache hits/misses of a single cache set while knowing the size of the backup cache.
Therefore, a dynamic backup cache resizing mechanism is proposed to mitigate this threat,
which dynamically resizes the backup cache based on the memory access count register
and sets enabled bits to 0 for cache lines to be disabled in the backup cache.

\subsection{Parallel Cache Access}
\begin{figure}[h!]
\centering
\includegraphics[width=\linewidth]{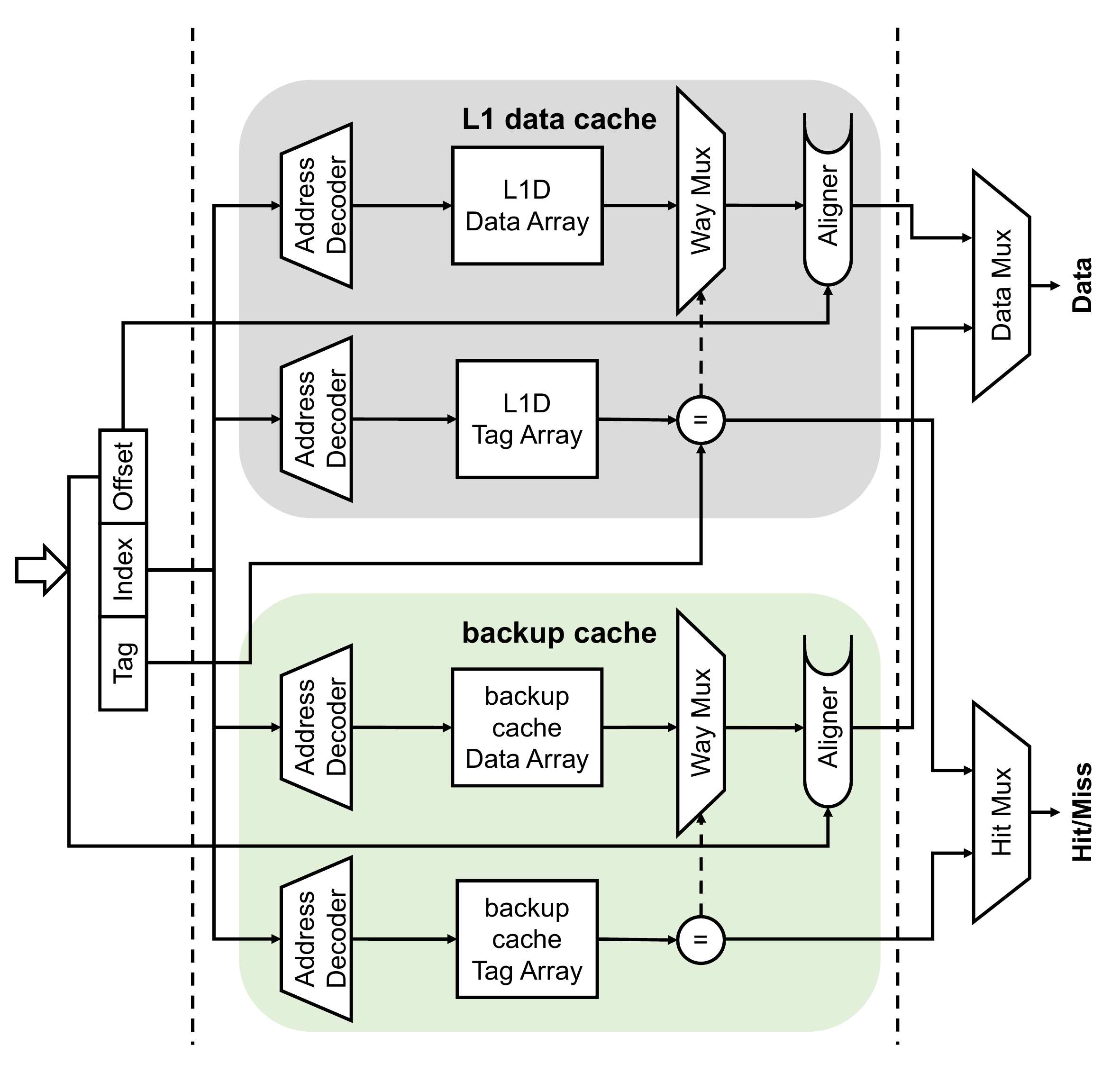}
\caption{Pipeline diagram for the L1 data cache and backup cache access.}
\label{fig-diagram}
\end{figure}

As the pipeline diagram for the L1 data cache and backup cache access shown in Figure~\ref{fig-diagram},
the L1 data cache access and the backup cache access are performed in parallel,
and thus the access latency of the L1 data cache and the backup cache
is the same with no additional leaks.
In addition, the L1 data cache hit or the backup cache hit results in the L1 cache hit,
while the L1 cache miss occurs only in the case of both the L1 data cache miss and the backup cache miss.

\begin{figure}[h!]
\centering
\includegraphics[width=\linewidth]{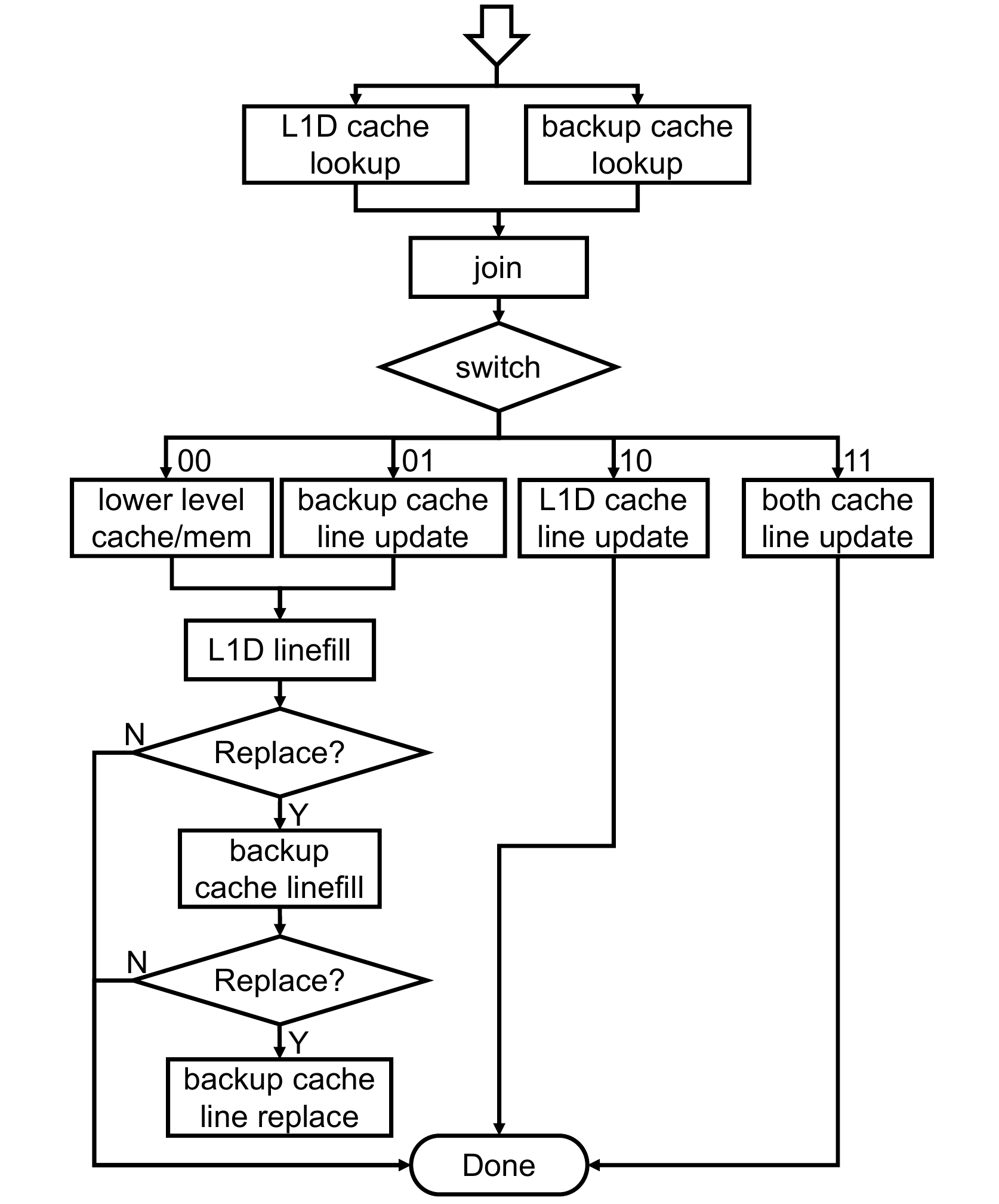}
\caption{The memory access process of L1 data cache and backup cache.}
\label{fig-access}
\end{figure}

Then, according to our microarchitecture design of BackCache,
there are four possible cache hit/miss scenarios for querying both the L1 data cache and the backup cache
when the processor performs memory access:
\begin{itemize}
\item 00: Both L1 data cache misses and backup cache misses.
\item 01: L1 data cache misses and backup cache hits.
\item 10: L1 data cache hits and backup cache misses.
\item 11: Both L1 data cache hits and backup cache hits.
\end{itemize}
Figure~\ref{fig-access} shows the process of L1 data cache and backup cache accesses
when the processor performs a memory access.
To avoid differences in access latency between these branches,
the line fills for both the L1 data cache from the backup cache
and the backup cache from the L1 data cache
are performed after the data has been responded to the processor,
similar to the write-back mechanism between multi-level caches.

The first case is when both the L1 data cache and the backup cache are cache misses,
meaning that the cache line does not exist in any L1-level cache.
Then, the L1 data cache forwards the memory access request to the lower-level cache or main memory,
and brings the data into the L1 data cache only.
And if the newly brought cache line results in a cache line replacement in the L1 data cache,
the evicted cache line is placed in the backup cache
with the used bit set to 0 based on the RURP replacement policy.
Besides, the backup cache only finds victims for replacement from cache lines with the enabled bit set to 1.

The second case happens when the L1 data cache misses and the backup cache hits,
which means that the data exists in the backup cache but not in the L1 data cache.
In this case, the data and the used bit are updated in the backup cache,
and this cache line is also brought into the L1 data cache at the same time.
Then, if this operation leads to a cache line replacement in the L1 data cache,
the backup cache handles the line fill operation in the same way as the first case.

The third case is handled in the same way as the original L1 data cache access,
where the L1 data cache hits and the backup cache misses,
meaning that the data only exists in the L1 data cache.
In this case, the processor only updates the data and relevant metadata in the L1 data cache,
and no changes are made to the backup cache.

The last case is that both the L1 data cache and the backup cache hit,
which means that the data exists in both the L1 data cache and the backup cache.
In this case, the processor updates the data and relevant metadata
in both the L1 data cache and the backup cache,
and the used bit is also set to 1 in the backup cache.

In addition, to maintain the cache coherency,
if a cache invalidation request is received from lower level cache,
the processor invalidates the corresponding cache line in both the L1 data cache and the backup cache.
And when a cache line is evicted from the L1 data cache,
a write-back request is sent to the lower-level cache or main memory if the dirty bit is set to 1,
no matter whether the cache line is placed in the backup cache or not.

\subsection{Random Used Replacement Policy (RURP)}
\begin{figure}
\centering
\includegraphics[width=1.7in]{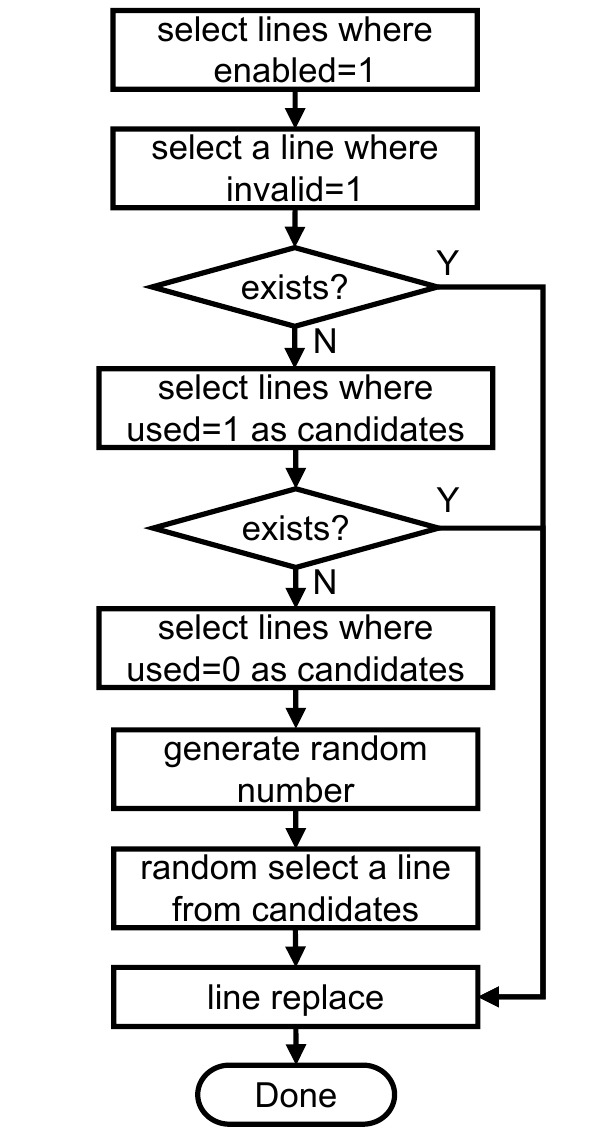}
\caption{The workflow of random used replacement policy (RURP).}
\label{fig-replacement}
\end{figure}

Random used replacement policy (RURP) for the backup cache is proposed to prevent
the cache lines that have not been reprobed by the attacker from being replaced in the backup cache.
As shown in Figure~\ref{fig-replacement}, the details of RURP are described as follows:
\begin{itemize}
\item When a cache line evicted from the L1 data cache is ready to be placed into the backup cache,
the backup cache first selects cache lines with the enabled bit set to 1.
\item Then, the backup cache selects an invalid cache line
(with the valid bit set to 0) as the victim cache line for replacement.
\item If there is no invalid cache line in the backup cache,
then the backup cache selects cache lines with the used bit set to 1 as candidates.
And if there is no cache line with the used bit set to 1,
then the backup cache selects cache lines with the used bit set to 0 as candidates.
\item Finally, the backup cache generates a random number and
then randomly selects a victim cache line from the candidate cache lines.
If there are no cache lines with the used bit set to 0,
the replacement is performed based on random selection.
The old cache line is replaced by the newly brought cache line.
\end{itemize}

In addition, to avoid eviction of cache lines in the backup cache
before they are re-accessed by the attacker,
the used bit is set to 0 during the context switch with the help of the operating system.
In this case, even if the attacker's cache line in the backup cache
is accessed by another process or by the same process in the first step of the attack,
the used bit is still set to 0 before the attacker re-probes this cache line.

\subsection{Dynamic Backup Cache Resizing Mechanism}
\begin{figure}
\centering
\subfloat[Resizing (init)]{\includegraphics[width=1.7in]{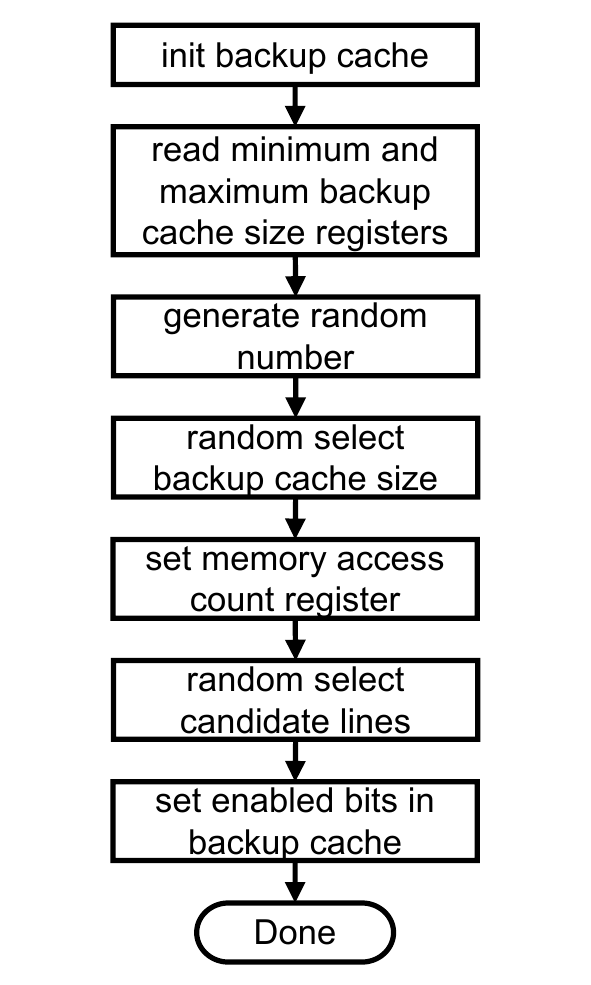}\label{fig-resize-initial}}
\subfloat[Resizing (resize)]{\includegraphics[width=1.7in]{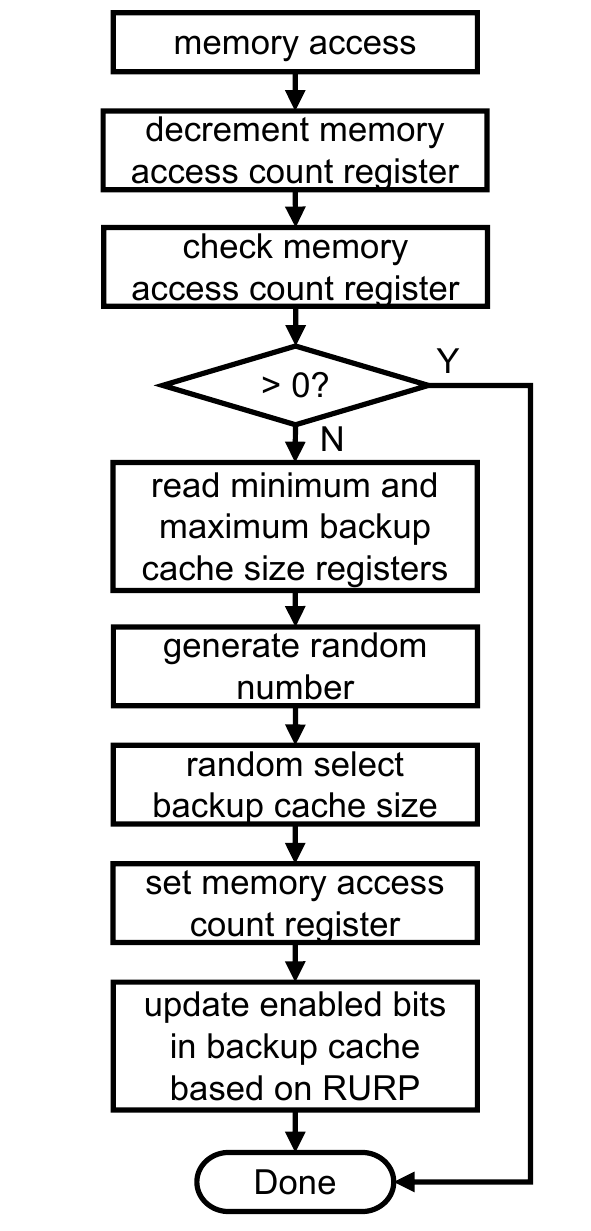}\label{fig-resize-resize}}
\caption{The workflow of dynamic backup cache resizing mechanism.}
\end{figure}

The dynamic backup cache resizing mechanism is proposed to
leave the attacker in the dark about the size of the backup cache.
As shown in Figure~\ref{fig-resize-initial}, when the cache is initialized,
the size of the backup cache is set to a random value
between the values of the minimum and maximum backup cache size registers,
and then the enabled bit for the same number of cache lines as the backup cache size is set to 1.
The value of the memory access count register is set to the same as the current backup cache size,
and each memory access operation decrements the value of this register by 1.

As Figure~\ref{fig-resize-resize} shows,
When the value of the memory access count register is decremented to 0,
the size of the backup cache is resized again to a random value between
the values of the minimum and maximum backup cache size registers,
and the value of the memory access count register is reset to this random value.
Then, if the size of the backup cache is not changed,
there is no need to change the enabled bit for the cache lines in the backup cache.
Besides, if the new size of the backup cache is larger than the old size of the backup cache,
the newly added cache lines are selected from the cache lines with the enabled bit set to 0,
and their enabled bit is set to 1.
Moreover, if the new size of the backup cache is smaller than the old size of the backup cache,
the backup cache selects cache lines according to the random used replacement policy (RURP)
to find the victim cache lines for replacement and set their enabled bit to 0.

\section{ISA and Software Support}
In this section, we describe the instruction set extensions and operating system changes to support BackCache.

One key aspect of our design relies on a used bit for random used replacement policy (RURP) to improve
the chances of keeping cache lines in the backup cache that are potentially not reprobed by the attacker.
During context switching, we need to reset the used bit of all cache lines.
To support this reset operation, we extend the instruction set architecture (ISA) with a new instruction:
\begin{itemize}
\item \textbf{\texttt{BUCLR}}:
The used bit of all cache lines in the backup cache should be cleared when the context is switched.
Thus, we add a BackCache Used Clear (\texttt{BUCLR}) instruction requiring no operands to support this feature.
This instruction is handled similarly to the Intel ISA's \texttt{CLFLUSH} instruction in the processor pipeline
since both instructions modify the tag bytes of the cache lines.
After executing the \texttt{BUCLR} instruction, the used bit of all cache lines in the backup cache are cleared.
In addition, if the design can clear the used bits of all entries in the tag array in parallel,
this instruction will take only a few cycles to complete.
The \texttt{BUCLR} instruction is a privileged instruction
and can only be executed in privileged mode (e.g. Linux kernel mode).
\end{itemize}

Then, we also modify the compiler to support the \texttt{BUCLR} instruction.
Our prototype implementation is based on the GNU GCC compiler and Binutils binary tools.
If the assembly code contains the \texttt{BUCLR} instruction,
the compiler will generate the corresponding machine code for the instruction.
Moreover, the operating system (OS) is responsible for switching between different contexts.
In our design, the OS needs to clear the used bit of all cache lines in the backup cache during context switching.
To support this feature, we modify the context switching code in the Linux kernel,
and insert the \texttt{BUCLR} instruction before switching to another process.

\section{Security Analysis}
In this section, we analyze the security of BackCache against contention-based cache timing attacks.
We make the following assumptions to analyze the security of BackCache:
\begin{itemize}
\item First, the attacker can construct the eviction set
of each cache set and the backup cache, and has enough time to infer
the victim's security-critical information based on the probing time of the eviction set.
\item Second, no other process accesses the target cache sets
during the alternate execution of the attacker and the victim.
This assumption is reasonable because if another process
other than the attacker or the victim evicts the attacker's cache lines,
the attack will be confused since it cannot know which cache set is accessed by the victim.
\item Third, the random generator used in the backup cache is secure and unpredictable.
\end{itemize}

\textbf{Security of Fully Associative Backup Cache.}
First, we discuss the security of the fully associative backup cache and the RURP policy.
Let $S$ and $W$ be the number of cache sets and the number of cache ways in the L1 data cache,
and the attacker can access these $S \times W$ cache lines to infer the victim's information.
Let $B$ be the number of cache lines in the backup cache,
and $V$ be the number of cache lines that can
evict an attacker's cache line in the L1 data cache during the victim's execution.
In the worst case, if the victim accesses $B+1$ cache lines (overflowing the backup cache),
the attacker can observe a cache miss to infer the victim's access pattern,
while the number is one for the baseline system.

For a 12-16KB dynamic backup cache, the worst-case number of $B+1$ is 193,
which means that the attacker can infer the victim's access pattern
if the victim accesses more than 12KB of data,
such as web browsers, video players, and so on.
Nevertheless, our defense still has a high probability of
preventing the attacker from obtaining timing information for several applications
that access less than 12KB of data in their time slice,
such as cryptographic applications (DES encryption, AES encryption, and so on).

Another interesting concern is whether the interaction between
the backup cache and the L1 data cache
causes observable side-channel leakage on higher-level caches (e.g. LLC).
First, if the cache line is evicted from the L1 data cache,
the state of the cache line in LLC is not affected.
Then, if the cache line is evicted from the backup cache,
the state of the cache line in the LLC is also unchanged.
Next, if a cache line is brought from the backup cache into the L1 data cache,
the state of the cache line in the LLC remains unaffected.
Thus, the interaction between the backup cache and the L1 data cache
does not cause any observable side-channel leakage to higher-level caches.

\textbf{Security of Dynamic Resizing Mechanism.}
Then, we analyze the security of the dynamic backup cache resizing mechanism.
Assume there is no backup cache overflow caused by the victim's execution.
Let $V$ be the number of cache lines that can evict an attacker's cache line
in the L1 data cache during the victim's execution.
Let $B_{min}$ and $B_{max}$ be the minimum and maximum cache lines in the backup cache,
$B$ be the number of enabled cache lines in the backup cache
when the attacker first fills the cache sets,
and $B^*$ be the number of enabled cache lines in the backup cache
when the attacker re-accesses the cache sets.

Even if the attacker knows the value of $B$ (in fact, the attacker does not know the value of $B$),
the indexes of cache sets accessed by the victim are still unknown.
Suppose that the attacker also fills potential victim cache sets
as well as $B$ cache lines in the backup cache, then:
\begin{itemize}
\item If $B^* \geq B+V$, the attacker will observe all cache hits
and will not be able to infer the victim's access information.
\item If $B^* < B+V$, the attacker will observe some cache misses in the eviction set,
but the fully associative backup cache makes the relationship between
these cache misses and the victim's access pattern unclear.
\end{itemize}
And if the attacker only fills potential victim cache sets
but does not fill any cache lines in the backup cache,
then the attacker will observe all cache hits during the re-accessing step,
without being able to distinguish the victim's access pattern.

Particularly, if the attacker only spies on a single cache set and knows the value of $B$, then:
\begin{itemize}
\item If $B^* > B$ ($B_{max}-B$ cases),
the attacker will observe cache hits for all cache lines in the eviction set,
but it is still possible that the victim accesses the cache set.
\item If $B^* = B$ (only one case),
the attacker will observe only one cache miss for cache lines filled into the backup cache,
and the attacker can obtain the correct access pattern of the victim in this case.
\item If $B^* < B$ ($B-B_{min}$ cases),
the attacker will always observe several cache misses for the eviction set,
yet the victim may not perform any memory access on the cache set.
\end{itemize}
The probability that the attacker can obtain the correct access pattern of the victim is:
\begin{equation}
\begin{aligned}
P(B) & = \frac{B_{max}-B}{B_{max}-B_{min}+1} \times p \\
& + \frac{1}{B_{max}-B_{min}+1} \\
& + \frac{B-B_{min}}{B_{max}-B_{min}+1} \times (1-p)
\end{aligned}
\end{equation}
Then, the average probability for every value of $B$ is:
\begin{equation}
\begin{aligned}
P_{avg} & = \frac{1}{B_{max}-B_{min}+1} \times \sum_{B=B_{min}}^{B_{max}} P(B) \\
& = \frac{1}{2} + \frac{1}{2(B_{max}-B_{min}+1)}
\end{aligned}
\end{equation}
The average probability is 0.5077, 0.5039, and 0.5026 for 12-16KB dynamic backup cache,
8-16KB dynamic backup cache, and 4-16KB dynamic backup cache.
And with $B_{max}-B_{min}+1$ getting larger and larger,
the attacker can only infer the victim's access pattern with a probability
that approximates a random guess (i.e., 0.5) even if the value of $B$ is known.

Since the probability of the attacker getting the
correct access pattern of the victim is not 0.5,
it is an interesting question to discuss
whether the attacker can repeat the attack to get the sensitive information.
And as our previous security evaluation experiment shows
(where we repeat the attack for different sizes of the eviction set),
the attacker is confused because the access time distribution is similar for both ‘0’ and ‘1’.

Another important issue is whether the dynamic resizing mechanism
will cause side-channel leakage at higher levels of the cache hierarchy (e.g., LLC).
Fortunately, the dynamic resizing mechanism only evicts cache lines
from the backup cache and does not evict cache lines from the L1 data cache,
so it will never cause the same cache line to be evicted from the LLC
for both inclusive and non-inclusive LLCs.

\textbf{Discussion.}
BackCache also provides certain security improvements over caches
without any defenses for attack strategies not covered in the threat model.
On the one hand,
while reuse-based cache timing attacks are outside the scope of this paper,
and BackCache does not break the sharing of cache lines in either temporal or spatial locality,
the fully associative backup cache does increase the complexity
(larger eviction set required) for the attacker to evict shared memory addresses,
making reuse-based attacks such as Evict+Reload~\cite{gruss2015cache} more difficult.
On the other hand,
if the prime operation on the backup cache
is performed after the victim's execution,
the attacker can still infer the sensitive cache set
through a four-step Prime+(Access)+Prime+Probe attack,
which is a complex attack strategy not covered in the threat model of this paper.
Nevertheless, for BackCache, the attacker must access at least 4 + 192 + 4 = 200
cache lines to determine whether a cache set is security-critical,
whereas for the baseline system,
they need to access only 4 + 4 + 4 = 12 cache lines.
This means BackCache increases the attacker's complexity by 16 times
compared to the baseline caches for a Prime+(Access)+Prime+Probe attack.

\section{Experimental Evaluation}
In this section, we first estimate the hardware overhead of our design using CACTI 6.5,
and then present the experimental setup on the gem5 simulator.
Subsequently, we evaluate the effectiveness of BackCache against
contention-based cache timing attacks with both lab and real-world attacks.
In addition, we also evaluate the performance overhead of BackCache
on context switching and user programs with different benchmarks.
Furthermore, we evaluate the performance impact and security guarantee
of the backup cache resizing frequency through
a sensitivity analysis of the memory access count.
Finally, we discuss and compare BackCache with existing cache defenses.

\subsection{Hardware Overhead}
We model our cache design using CACTI 6.5
to estimate the access latency and energy cost of our design, assuming a 32nm technology.
According to the simulation results listed in Table~\ref{tab-cacti},
the access latency of the backup cache is 1.54 times that of the L1 data cache
due to the additional circuitry of the fully associative cache.
Meanwhile, to guarantee that there is no side-channel risk in the access latency between these two caches,
the access latency of the BackCache increases from approximately 2 clock cycles to 3 clock cycles.
Then, since both the dynamic backup cache and the L1 data cache
are requested and accessed in a single memory access,
the energy consumption of BackCache is the sum of these two caches,
which increases to 1.77 times that of the baseline system.
Furthermore, we conducted an area overhead assessment of BackCache using CACTI 6.5.
The results indicate that the cache system's area increases by approximately 2.05 times.
Despite the noticeable area overhead,
it remains within acceptable limits for the overall processor chip area.

\begin{table}[h!]
\footnotesize
\centering
\caption{Access Latency and Energy Estimation using CACTI 6.5}
\label{tab-cacti}
\begin{tabular}{cccc}
\toprule
\textbf{Configuration} & \textbf{\makecell{Baseline \\ (32K 4w)}} &
\textbf{\makecell{BackCache \\ (16K 4w+16K full)}} & \textbf{Overhead} \\
\midrule
L1I Time ($ns$) & 0.2642 & 0.2642 & --- \\
L1D Time ($ns$) & 0.2925 & 0.4507 & 1.54x \\
L2 Time ($ns$) & 2.8783 & 2.8783 & --- \\
L1D Energy ($nJ$) & 0.0325 & 0.0577 & 1.77x \\
\bottomrule
\end{tabular}
\end{table}

\subsection{Experimental Setup on Gem5}
\textbf{Experimental Settings.}
We implemented BackCache in the gem5 cycle-accurate simulator,
using a 32KB 2-way set-associative L1 instruction cache per core,
a 16KB 4-way set-associative L1 data cache
(with a 16KB dynamic fully associative backup cache) per core,
and a 1MB 8-way set-associative L2 (LLC) cache shared across all cores.
We add a used bit and an enabled bit to each cache line in the backup cache
and three hardware registers to the processor core.
Based on our hardware overhead estimation using CACTI 6.5,
the access latency of the L1 instruction cache is set to 2 cycles for both systems,
and the access latency of the L2 cache is set to 20 cycles for both systems.
Then, the access latency of the L1 data cache is set to 2 cycles for the baseline system,
and set to 3 cycles for the BackCache system.
Besides, the DRAM is modeled as a single-channel DDR3-1600 (8x8 topology) memory controller.
Moreover, the simulator runs a 64-bit version of Ubuntu 18.04
with the Linux kernel 4.19.264 when evaluating in full system mode.
The simulator configuration of the baseline system (without any defenses)
and BackCache system are shown in Table~\ref{tab-config}.
In addition, to evaluate the impact of the dynamic backup cache size on system performance,
we configure the 16KB backup cache with three different parameters:
\begin{itemize}
\item BackCache (12-16KB): the minimum and maximum sizes of the backup cache are 12KB and 16KB.
\item BackCache (8-16KB): the minimum and maximum sizes of the backup cache are 8KB and 16KB.
\item BackCache (4-16KB): the minimum and maximum sizes of the backup cache are 4KB and 16KB.
\end{itemize}

\begin{table}[h!]
\footnotesize
\centering
\setlength{\tabcolsep}{2.1pt}
\caption{Simulator Configuration of Baseline and BackCache}
\label{tab-config}
\begin{tabular}{ccc}
\toprule
\textbf{Parameter} & \textbf{Baseline} & \textbf{BackCache} \\
\midrule
ISA & AArch64 & AArch64 \\
CPU & DerivO3CPU, 2.0GHz & DerivO3CPU, 2.0GHz \\
L1I & 2-way 32KB, 2 cycles, LRU & 2-way 32KB, 2 cycles, LRU \\
L1D & 4-way 32KB, 2 cycles, LRU & 4-way 16KB, 3 cycles, LRU \\
Backup & None & Dynamic 16KB, 3 cycles, RURP \\
L2 (LLC) & 8-way 1MB, 20 cycles, LRU & 8-way 1MB, 20 cycles, LRU \\
DRAM & DDR3-1600, 8x8, 16GB & DDR3-1600, 8x8, 16GB \\
\bottomrule
\end{tabular}
\end{table}

\textbf{Security Evaluation Workloads.}
For security evaluation, we create Prime+Probe attacks targeting
one cache set and an OpenSSL version of AES encryption.
In the first case, the attacker first primes the potential secret cache set and the backup cache,
then the victim accesses this cache set if the secret is 1,
and then the attacker probes the potential secret cache set
to extract information about the secret based on the access time.
In the second case, the attacker also primes the potential secret cache sets first,
then the victim accesses the memory location of the T-tables,
and then the attacker probes the potential secret cache sets
to extract information about indexes of the T-tables being used based on the access time.

\textbf{Performance Evaluation Workloads.}
We validate the performance of BackCache on both Linux kernel and user programs.
We modify the Linux kernel 4.19.264 by adding a \texttt{BUCLR} instruction
during context switching for implementing BackCache,
while the baseline system runs the original Linux kernel 4.19.264.
For Linux kernel evaluation, we use lat\_ctx in LMBench 3.0 to
evaluate context switch overhead in full system mode.
For user program performance evaluation,
we use the Mibench benchmark suite with “large” input sets,
SPEC 2017 benchmark suite with “ref” input sets,
and PARSEC 3.0 benchmark suite with “simlarge” input sets.

\textbf{Evaluation Metrics.}
We use the following metrics to evaluate the effectiveness and performance of BackCache:
\begin{itemize}
\item \textbf{Access latency}: Access latency when an attacker re-accesses the primed cache sets
is used to evaluate the security of BackCache,
and the attacker cannot observe the time difference between
the eviction and non-eviction of the cache set.
\item \textbf{Context switch latency}: For Linux kernel evaluation,
we use the context switch latency to measure the OS overhead,
i.e., the time to save the state of the current process and restore the state of the next process.
\item \textbf{IPC}: We use the number of instructions per cycle (IPC)
as the metrics to evaluate the performance of single-thread user programs.
\item \textbf{ROI}: We use the region of interest (ROI) instead of IPC to
evaluate the performance of multi-thread user programs,
which is an accepted metric for evaluating multi-thread performance.
\end{itemize}

\subsection{Security Evaluation}

\textbf{Case Study I: Single Cache Set Attack.}
In this case, the attacker knows the cache set that the victim accesses,
and the victim accesses only one cache set during execution.
Since the Prime+Probe attack is a typical contention-based cache timing attack,
we use it to evaluate the security of BackCache.

The attacker first primes the cache set (including the backup cache if exists) with the eviction set,
then the victim accesses this cache set based on the value of the secret.
When the secret data is 0, the victim does not access the cache,
and the attacker observes a cache hit when re-accessing the cache set.
When the secret data is 1, the victim accesses this cache set,
causing the eviction of the cache lines primed by the attacker,
resulting in a cache miss when the attacker re-accesses the cache set.

\begin{figure}[h!]
\centering
\subfloat[Baseline]{\includegraphics[width=0.5\linewidth]{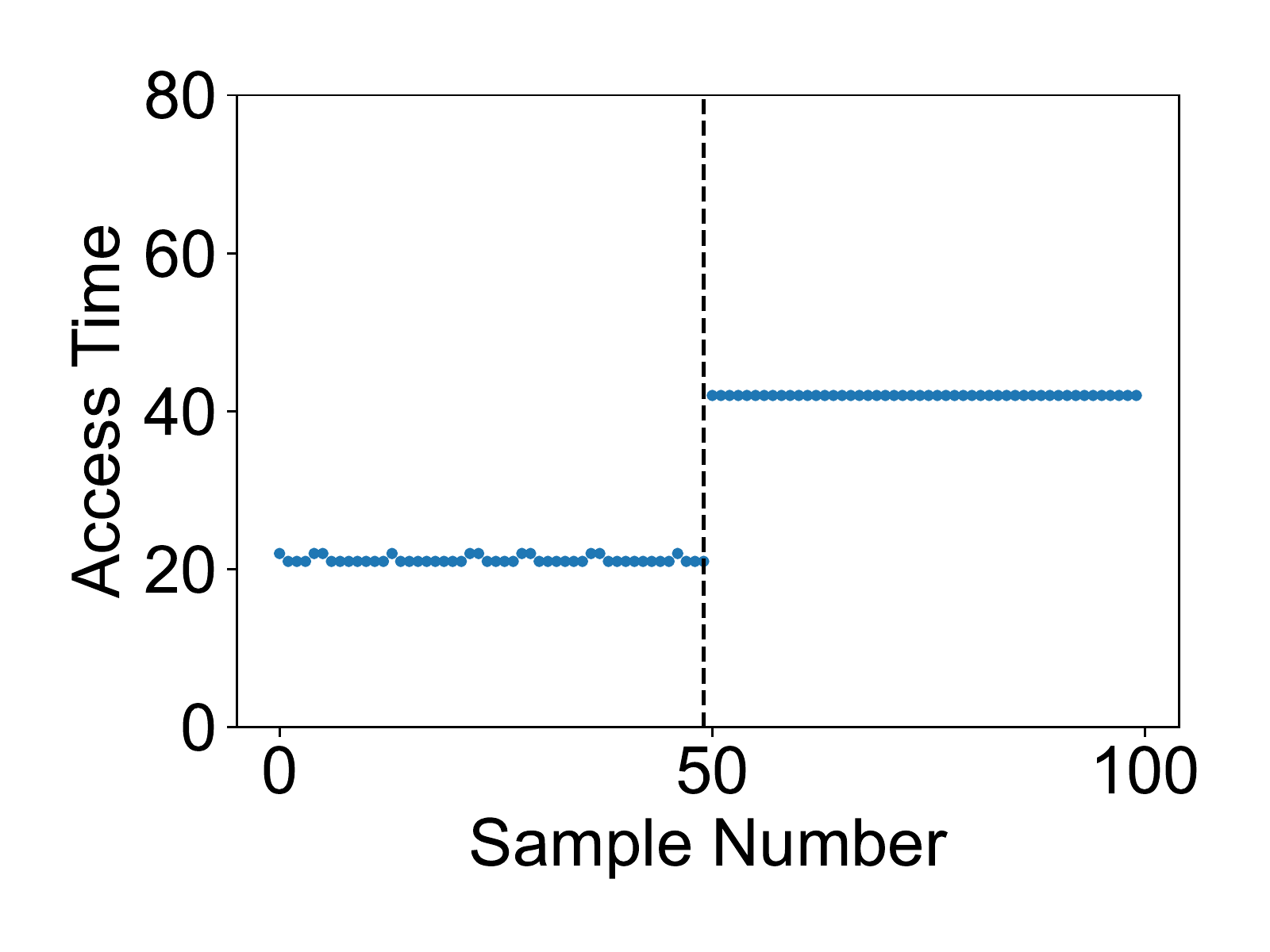}}
\subfloat[BackCache (0KB eviction)]{\includegraphics[width=0.5\linewidth]{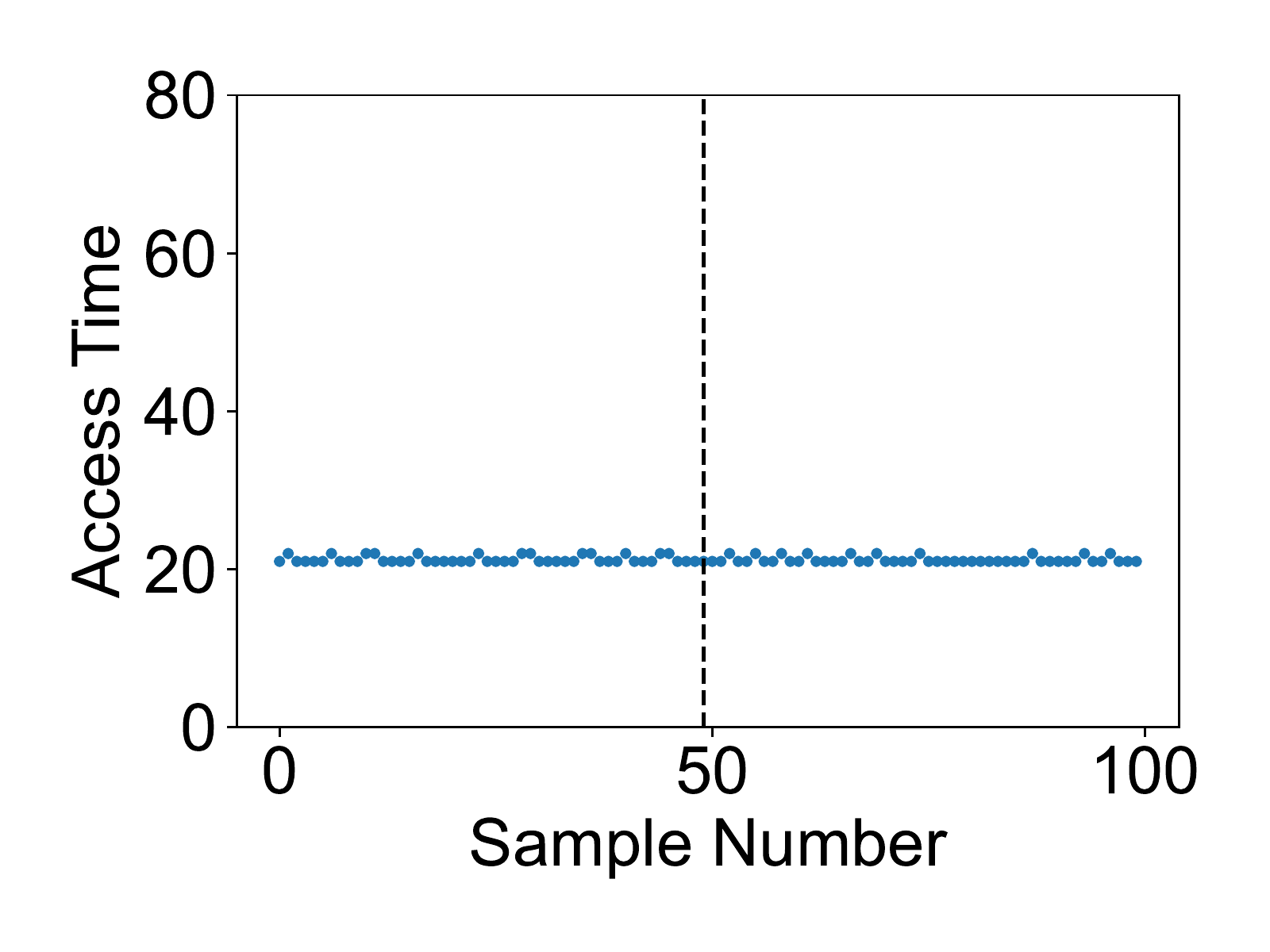}}\\
\subfloat[BackCache (4KB eviction)]{\includegraphics[width=0.5\linewidth]{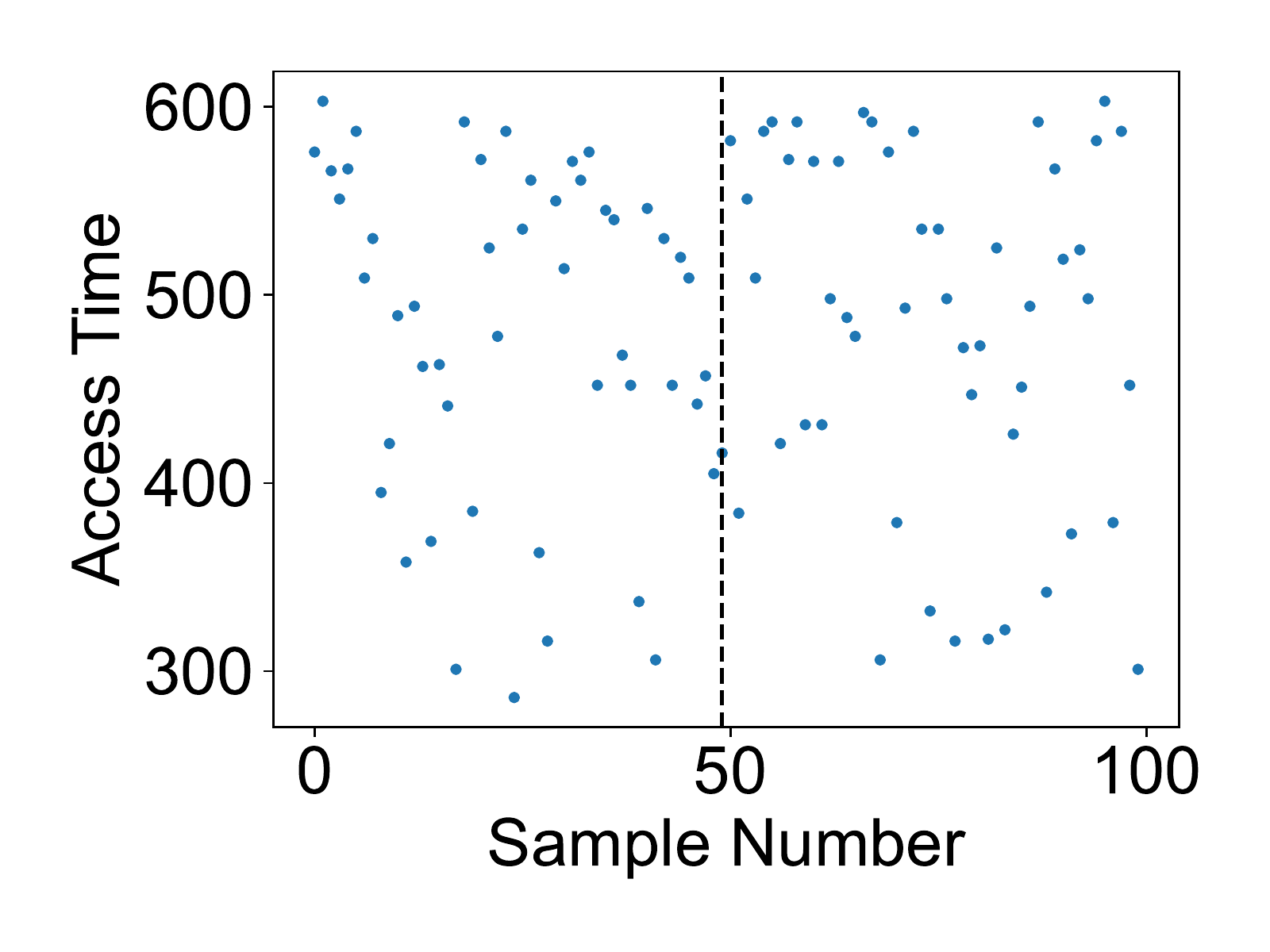}}
\subfloat[BackCache (8KB eviction)]{\includegraphics[width=0.5\linewidth]{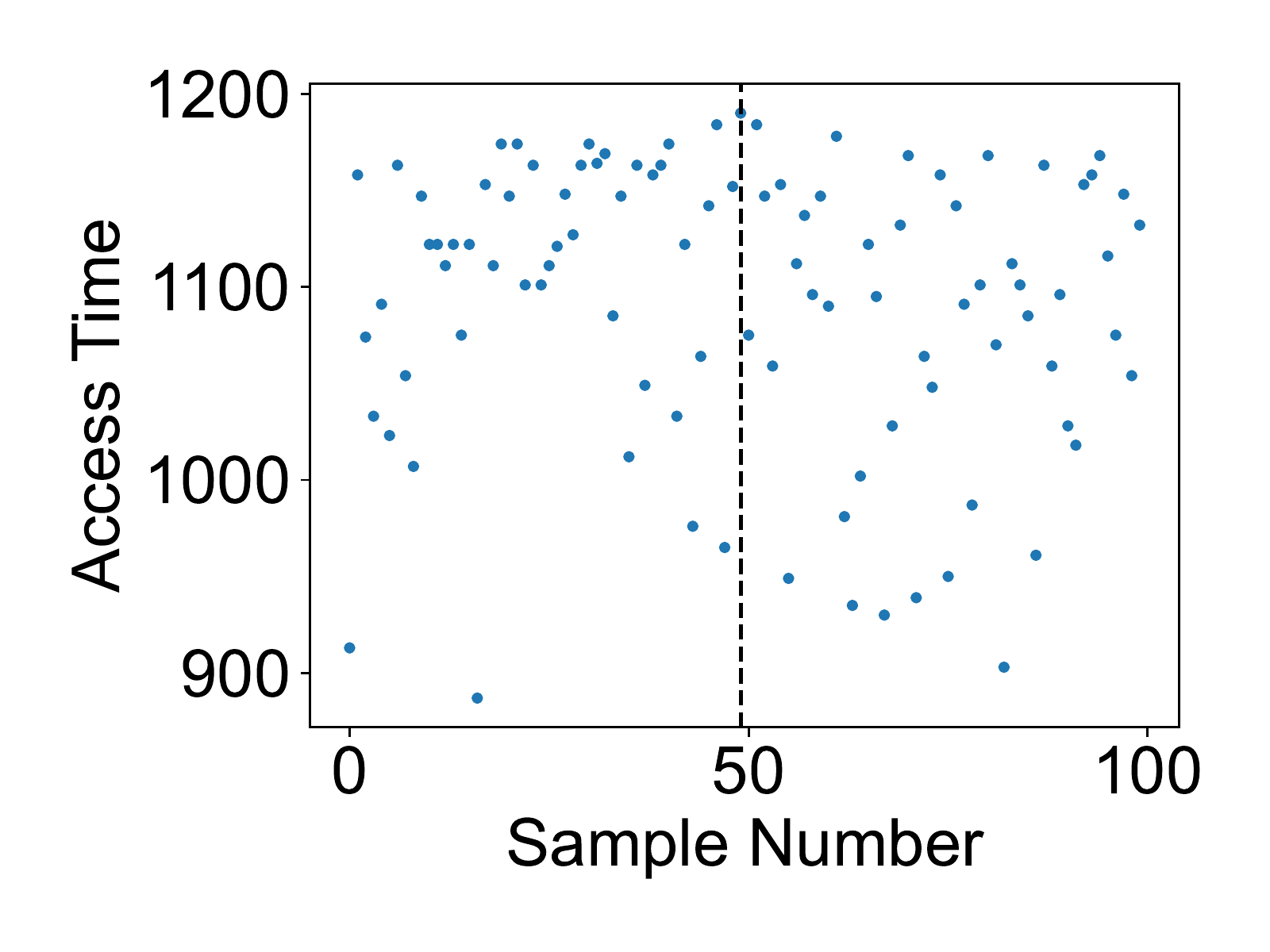}}\\
\subfloat[BackCache (12KB eviction)]{\includegraphics[width=0.5\linewidth]{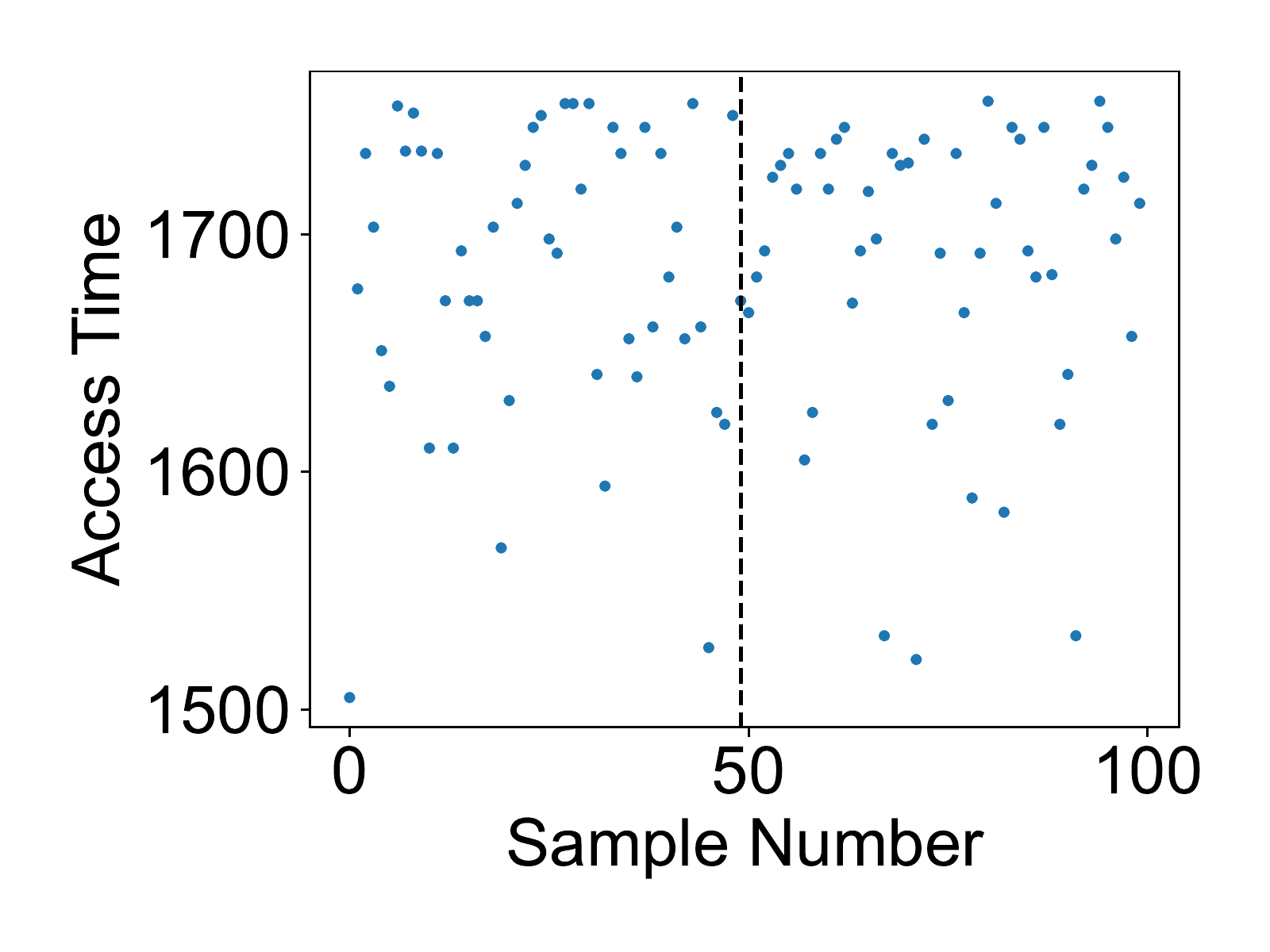}}
\subfloat[BackCache (16KB eviction)]{\includegraphics[width=0.5\linewidth]{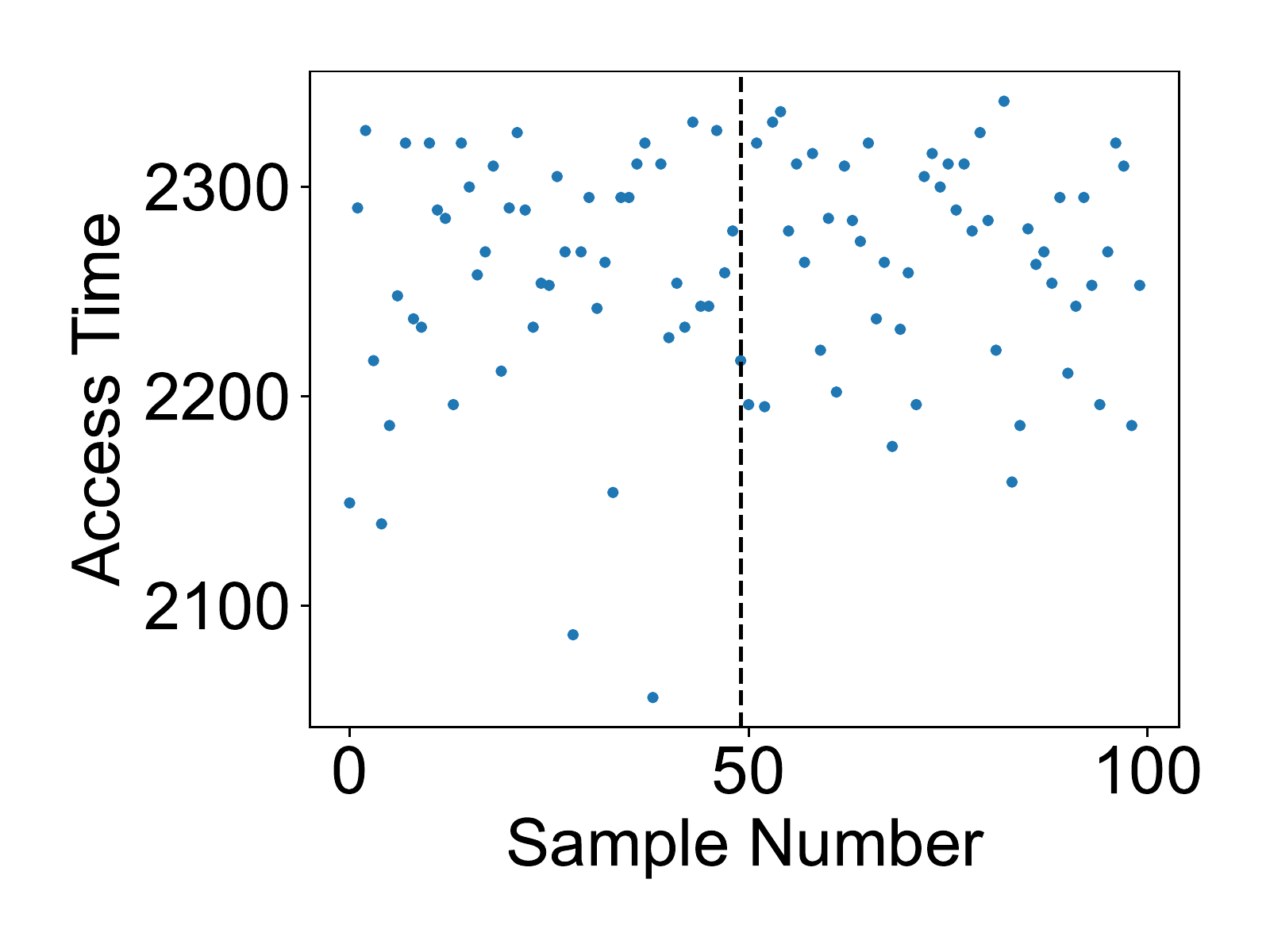}}
\caption{Evaluation results of the Prime+Probe attack on the single cache set attack.
The access time represents the time to access the entire eviction set,
including both the L1 data cache and the backup cache.}
\label{fig-single}
\end{figure}

Figure~\ref{fig-single} reports the evaluation result of the baseline system
and BackCache system with 12-16KB dynamic backup cache against the single cache set attack.
The attacker extracts 100 secret bits from the victim,
while the first 50 bits are 0 and the last 50 bits are 1.
The attacker primes the backup cache with a 4KB eviction set, 8KB eviction set,
12KB eviction set and 16KB eviction set for this case
while filling the original cache set with the proper eviction set.

The results show that the baseline system is vulnerable to attacks on a single cache set,
while the BackCache system protects the secret bits from being extracted by an attacker,
regardless of how large the eviction set is.
For example, when an attacker only fills the original cache set,
then all cache hits are observed when the cache set is re-accessed
because the backup cache handles the eviction of the cache set.
In addition, due to the dynamic backup cache resizing,
an attacker cannot observe any valid information about the victim's access pattern,
even if the attacker primes the backup cache with a large eviction set.

\textbf{Case Study II: Real-World Multiple Cache Sets Attack.}
In this case, the attacker primes the cache sets that may be used for the T-tables
of AES encryption (implemented in OpenSSL 1.1.1m) in the victim process with the eviction set.
Then after the victim's memory access, the attacker probes the potential secret cache sets
to infer the activity of encryption.

\begin{figure}[h!]
\centering
\subfloat[Baseline]{\includegraphics[width=0.5\linewidth]{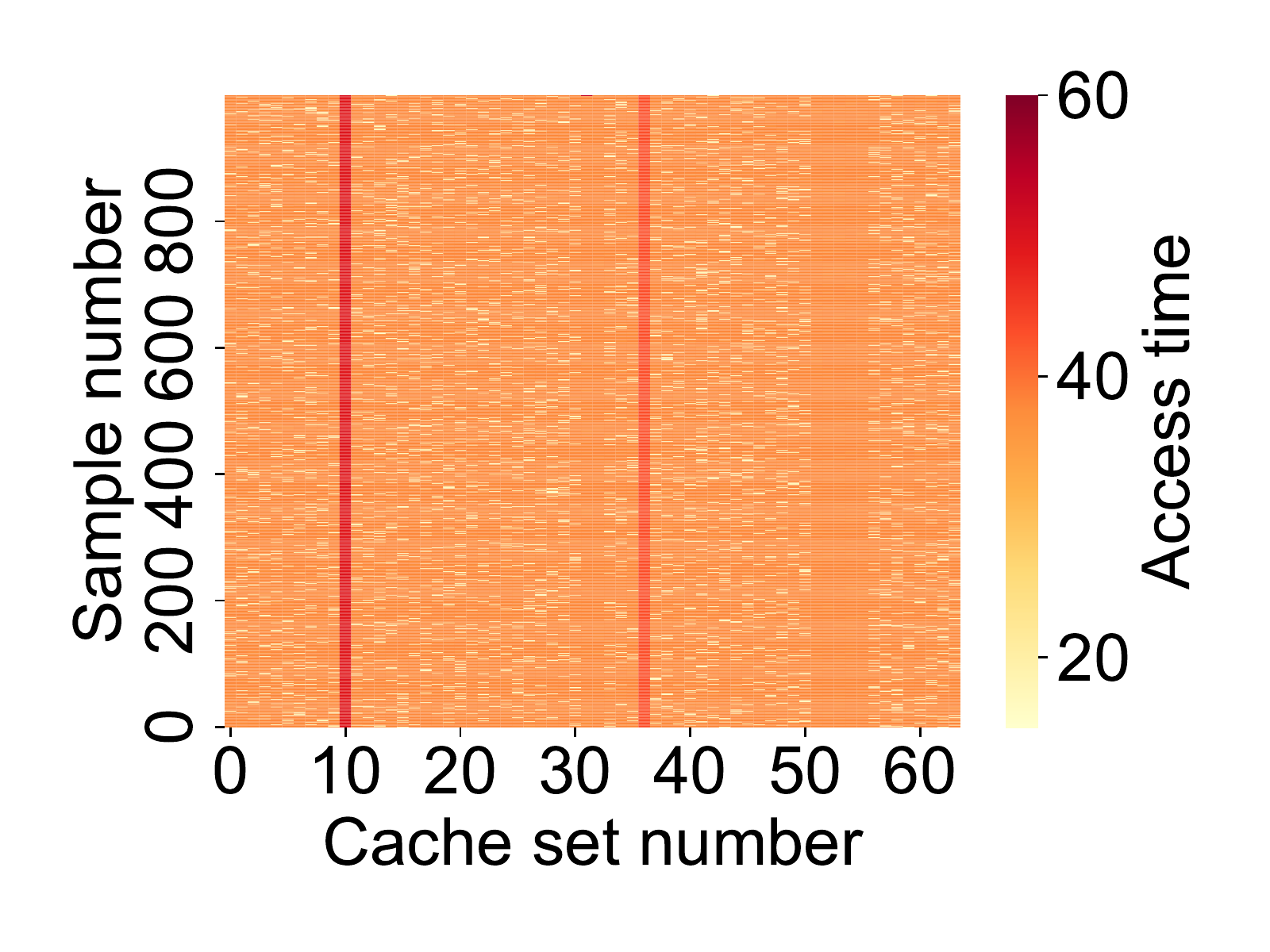}}
\subfloat[BackCache (12-16KB)]{\includegraphics[width=0.5\linewidth]{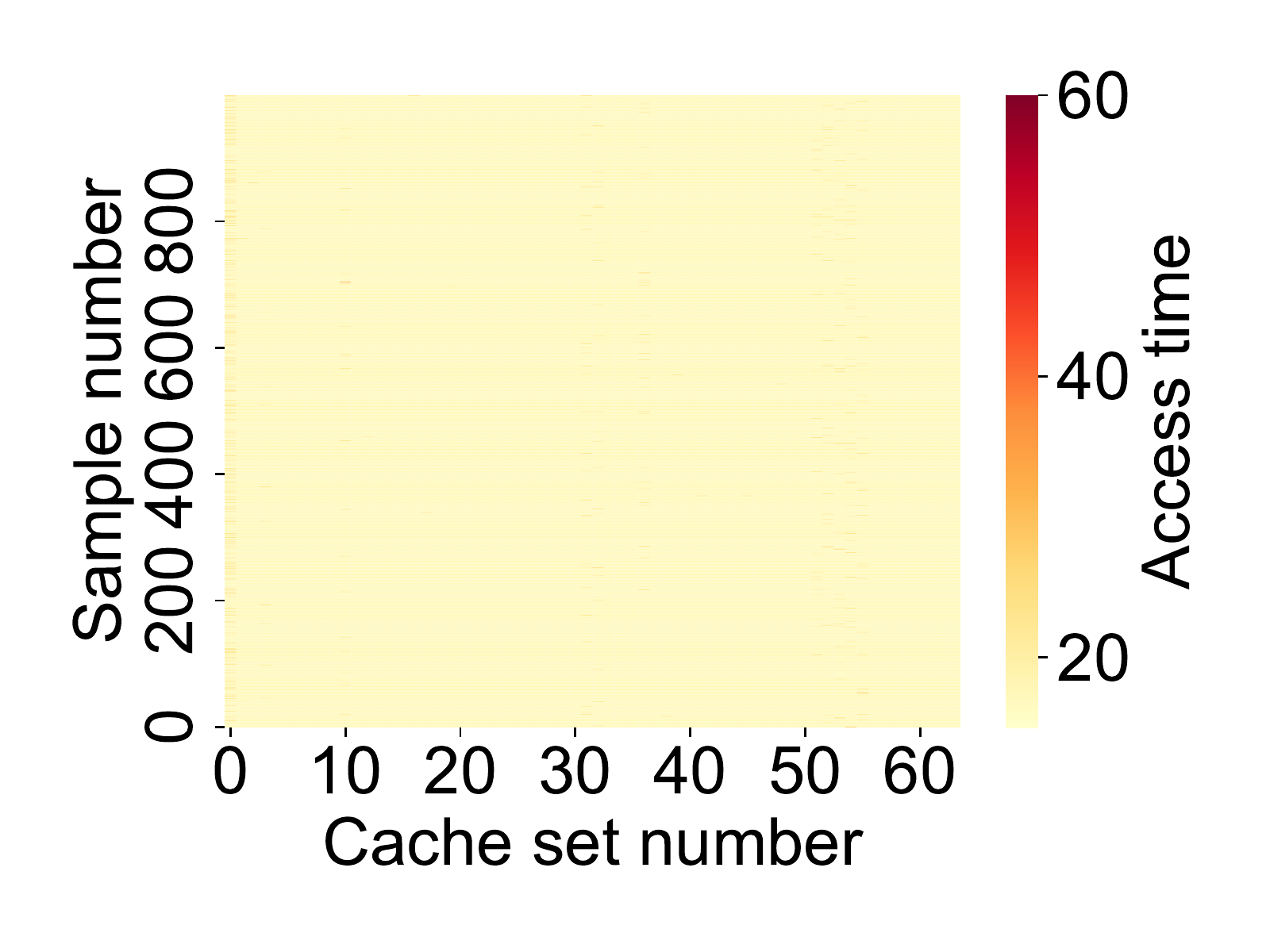}}\\
\subfloat[BackCache (8-16KB)]{\includegraphics[width=0.5\linewidth]{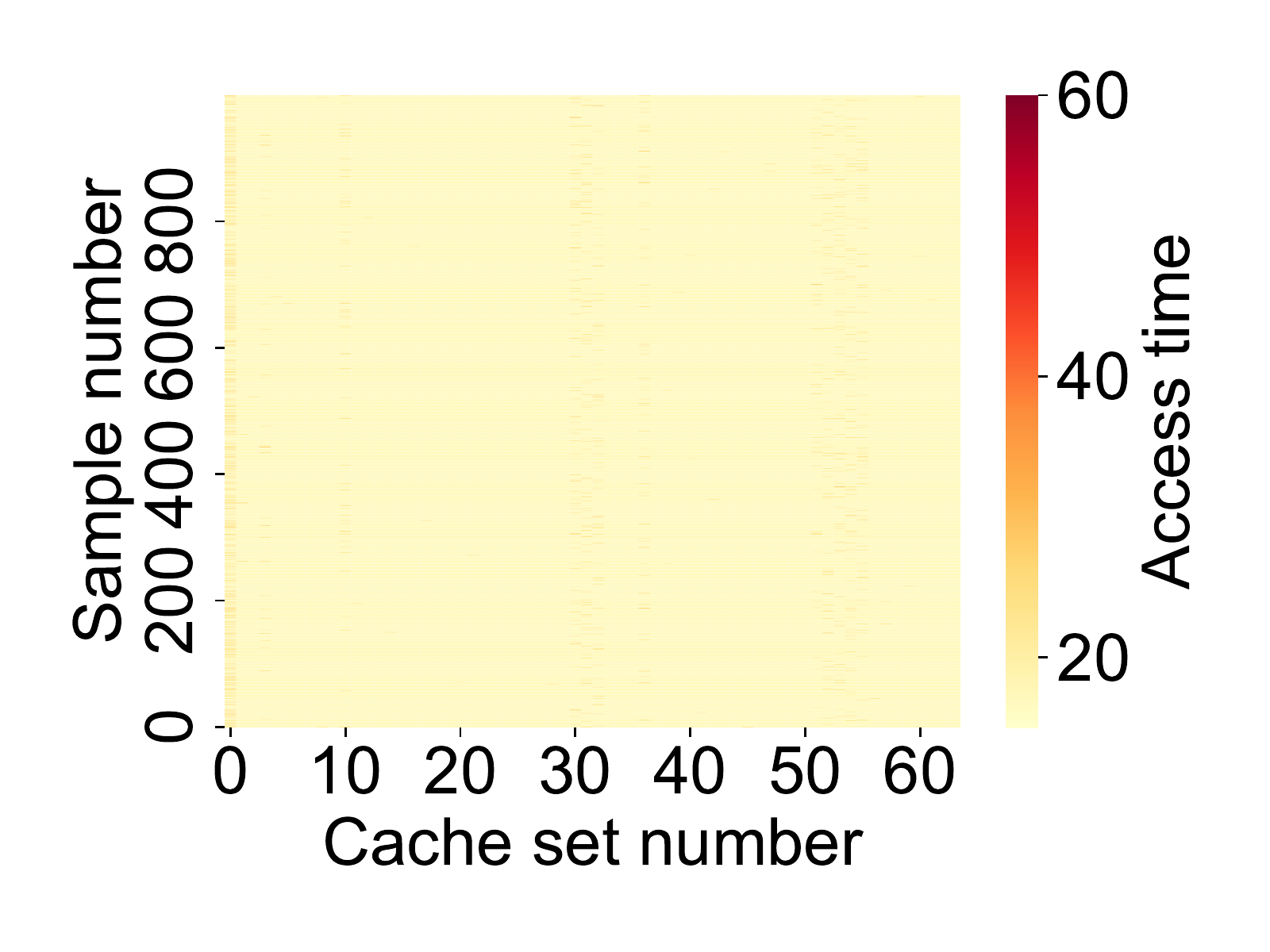}}
\subfloat[BackCache (4-16KB)]{\includegraphics[width=0.5\linewidth]{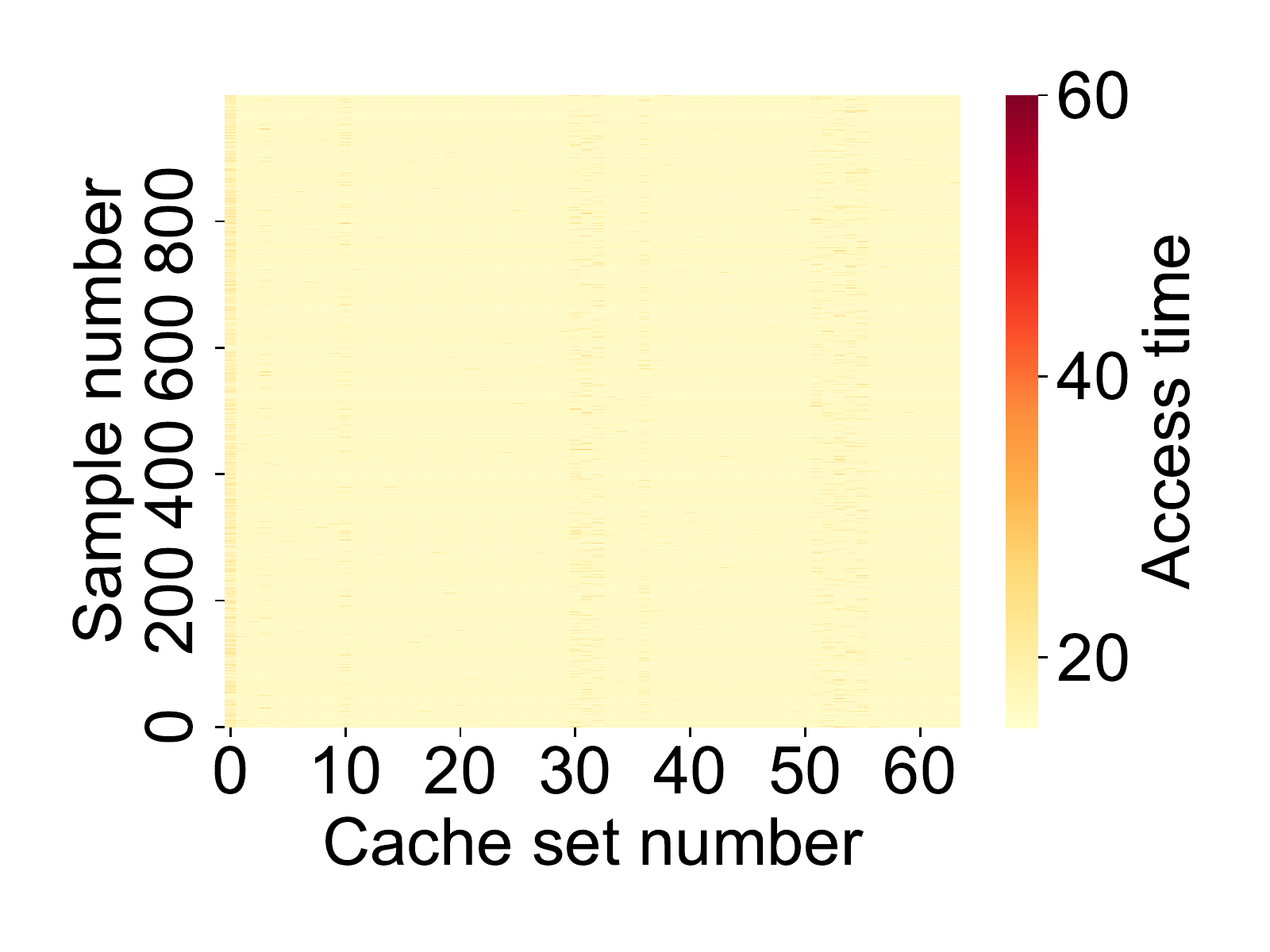}}
\caption{Evaluation results of the Prime+Probe attack on the AES encryption implemented in OpenSSL 1.1.1m.}
\label{fig-aes}
\end{figure}

Figure~\ref{fig-aes} depicts the results of the unprotected baseline system and BackCache system
proposed in this paper against the Prime+Probe attack
on OpenSSL version of AES encryption in a heat map.
The x-axis represents the cache set number from 0 to 63
since the AES-encrypted T-table occupies 64 different cache lines located in 64 different cache sets.
The y-axis is the number of samples, with a total of 1000 samples,
and each sample represents a group of re-accesses to the cache sets.
Then the color represents the access latency of the re-accesses,
and a stronger red color means a higher access latency.

We can see that the unprotected baseline system is vulnerable to the attack,
since the attacker can observe the cache misses for some cache sets
(clear color difference between cache sets) due to the victim's memory accesses.
On the contrary, our BackCache system is not vulnerable to the attack,
while the attacker obtains cache hits for all cache sets
(similar color across cache sets) when re-accessing them.
Thus, the attacker cannot learn whether the victim is accessing the T-tables
and then cannot extract enough information to break the AES encryption.
This result also demonstrates that BackCache can
prevent the contention-based cache timing attacks under real-world scenarios.

\subsection{Kernel Overhead Evaluation}
\begin{figure*}[h!]
\centering
\subfloat[BackCache (12-16KB)]{\includegraphics[width=0.33\linewidth]{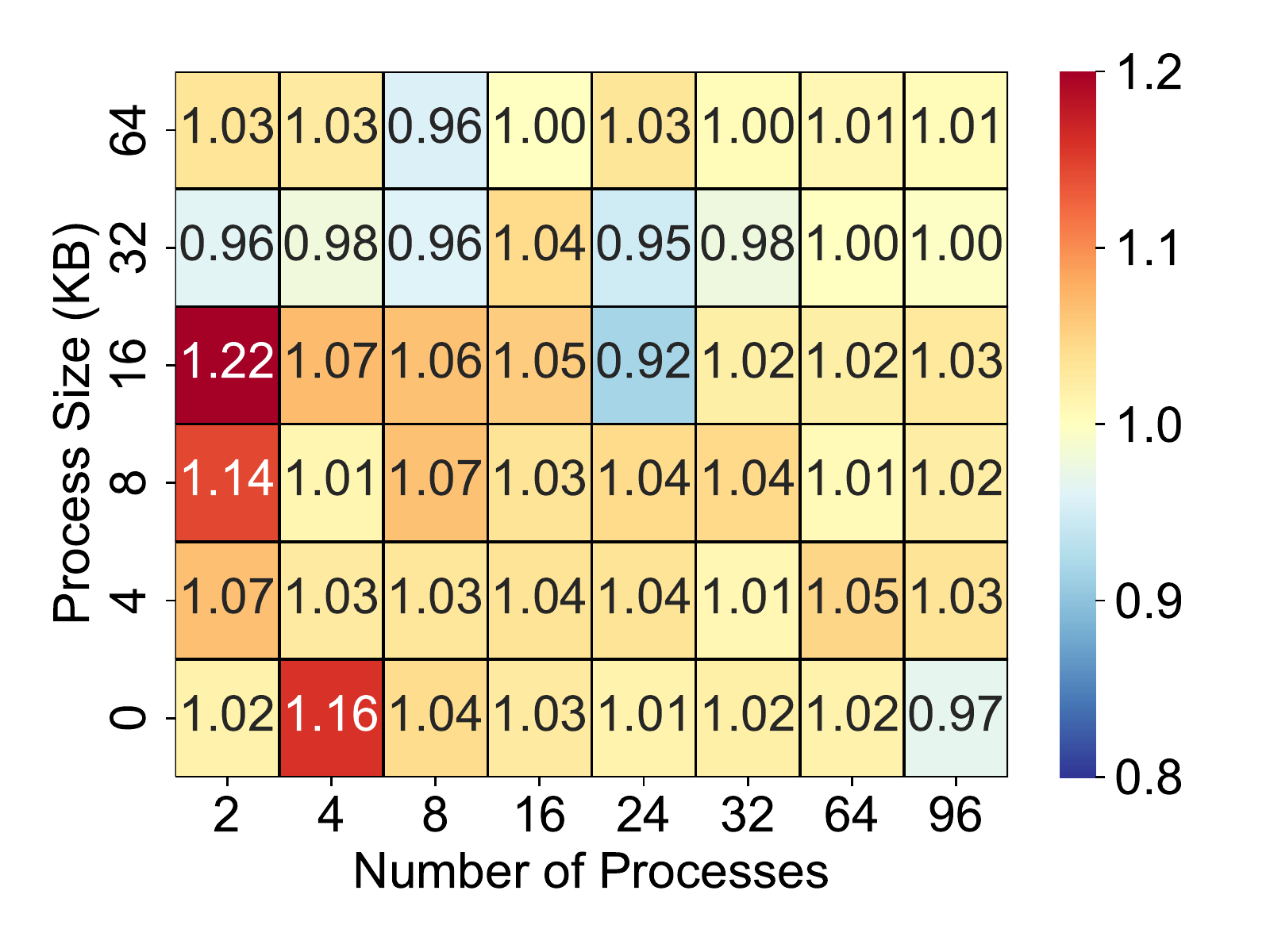}}
\subfloat[BackCache (8-16KB)]{\includegraphics[width=0.33\linewidth]{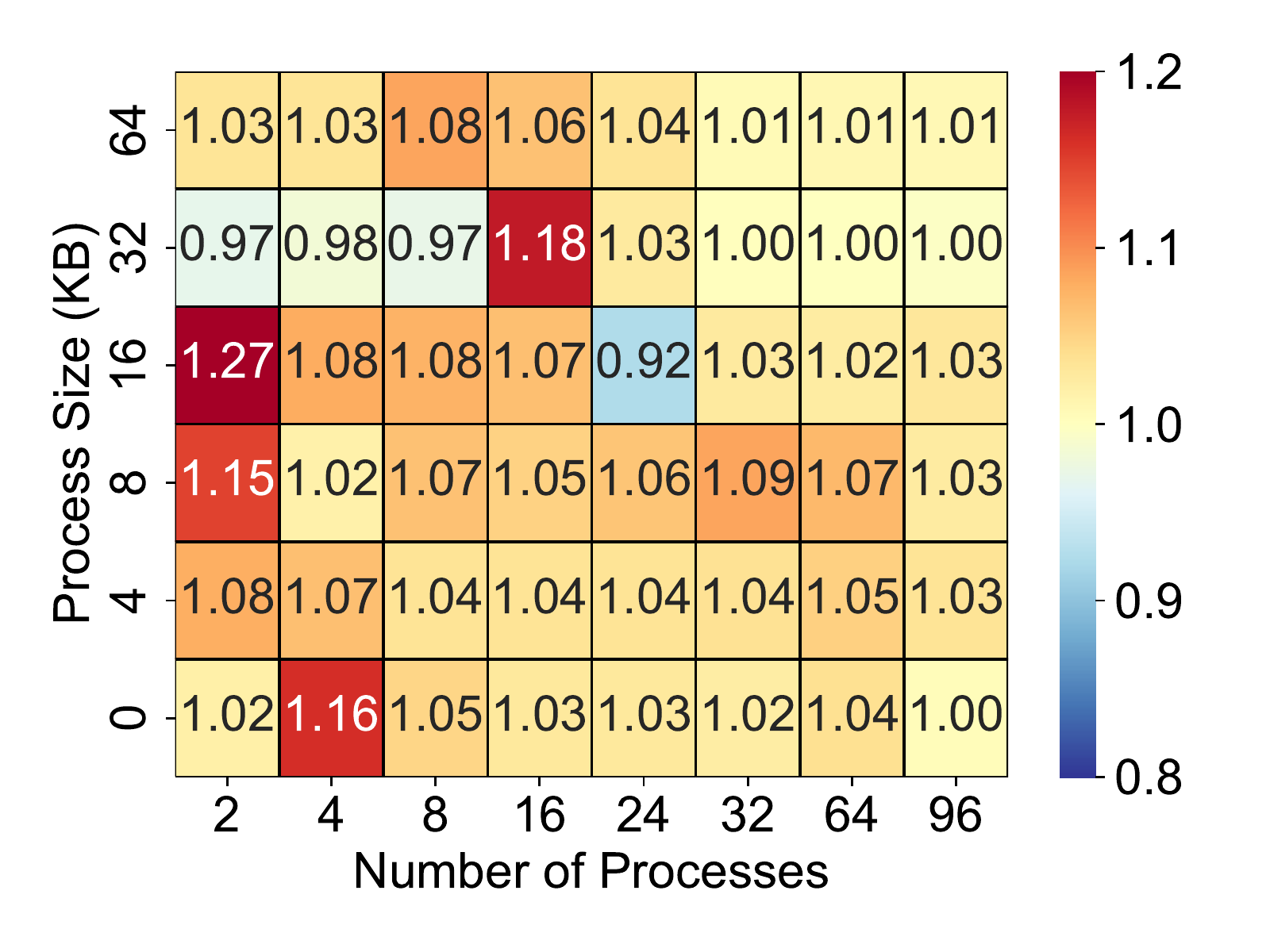}}
\subfloat[BackCache (4-16KB)]{\includegraphics[width=0.33\linewidth]{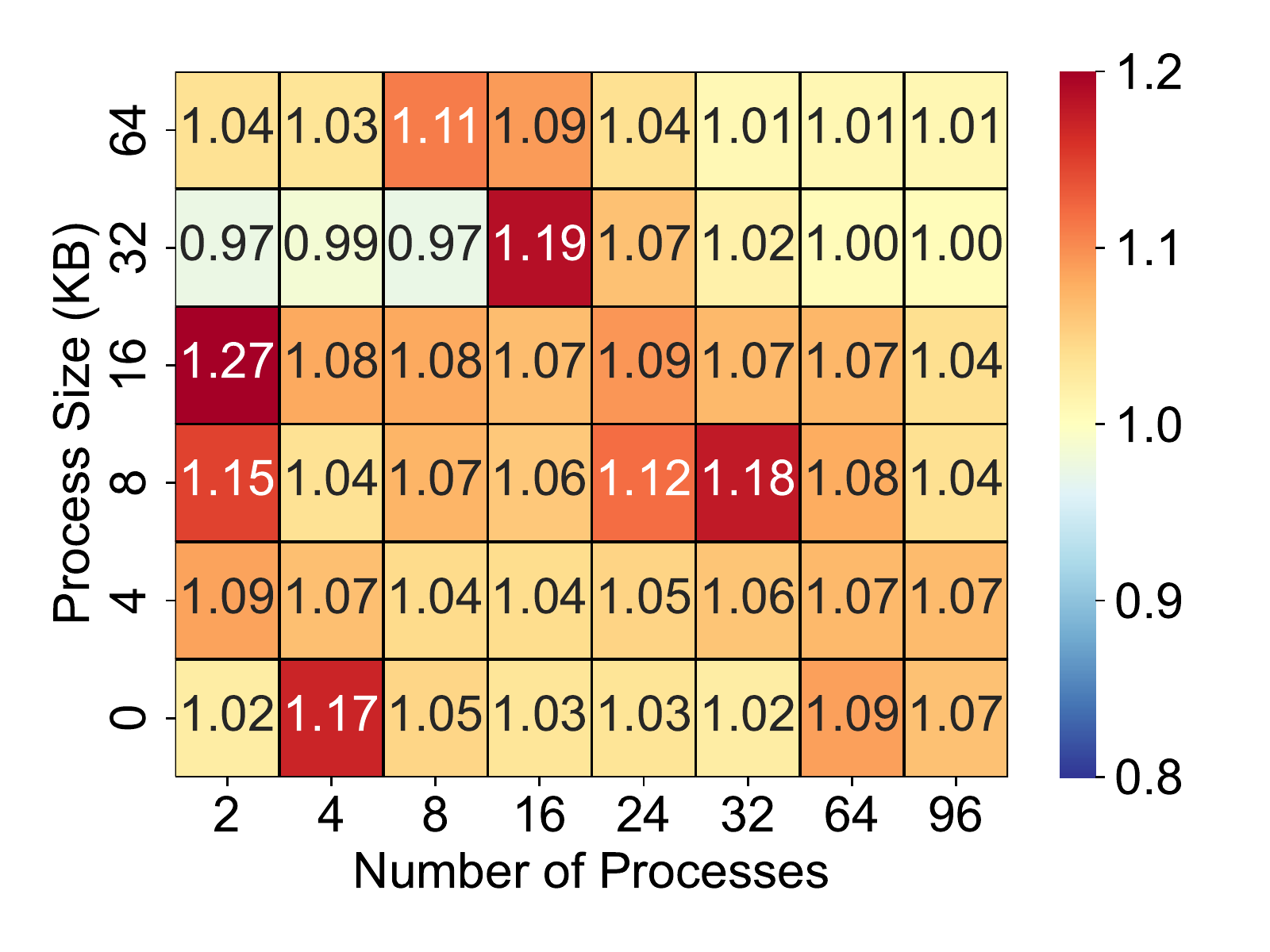}}
\caption{Context switch latency comparison between the baseline and BackCache systems.}
\label{fig-context-switch}
\end{figure*}

Table~\ref{tab-context-switch} and Figure~\ref{fig-context-switch}
show the evaluation result of context switch latency for
the baseline and BackCache system with different process sizes and numbers,
and the different dynamic backup cache sizes.
The experimental results show that the overhead of context switching increases by 2.61\% on average
for BackCache with 12-16KB dynamic backup cache compared to the baseline system,
with a 8\% reduction in latency in the best case
(from 4.34µs to 3.98µs when the process size is 16KB and the process number is 24)
and a 22\% increase in latency in the worst case
(from 2.73µs to 3.33µs when the process size is 16KB and the process number is 2).
Then, the average context-switch latency overhead for BackCache with 8-16KB dynamic backup cache is 4.47\%,
where the latency is reduced by 8\% in the best case
(from 4.34µs to 4.01µs when the process size is 16KB and the process number is 24)
and increased by 27\% in the worst case
(from 2.73µs to 3.47µs when the process size is 16KB and the process number is 2).
In addition, BackCache with 4-16KB dynamic backup cache
results in an average context switch latency overhead of 6.33\%,
with a 3\% reduction in latency in the best case
(from 3.75µs to 3.65µs when the process size is 32KB and the process number is 8)
and a 27\% increase in latency in the worst case
(from 2.73µs to 3.47µs when the process size is 16KB and the process number is 2).

\begin{table}[h!]
\footnotesize
\setlength{\tabcolsep}{3.2pt}
\centering
\caption{Context Switch Latency for Baseline and BackCache with Different Process Sizes and Numbers}
\label{tab-context-switch}
\begin{tabular}{cccccccccc}
\toprule
\textbf{System} & \textbf{Size} & \textbf{2} & \textbf{4} &
\textbf{8} & \textbf{16} & \textbf{24} & \textbf{32} &
\textbf{64} & \textbf{96} \\
\midrule
Baseline & \makecell{0KB \\ 4KB \\ 8KB \\ 16KB \\ 32KB \\ 64KB} &
\makecell{3.17 \\ 3.32 \\ 3.33 \\ 2.73 \\ 3.02 \\ 2.74} &
\makecell{3.35 \\ 3.92 \\ 4.07 \\ 4.00 \\ 4.22 \\ 3.96} &
\makecell{3.40 \\ 3.68 \\ 3.67 \\ 3.54 \\ 3.75 \\ 3.79} &
\makecell{3.47 \\ 3.67 \\ 3.74 \\ 3.64 \\ 4.18 \\ 16.95} &
\makecell{3.50 \\ 3.69 \\ 3.76 \\ 4.34 \\ 8.44 \\ 23.43} &
\makecell{3.51 \\ 3.76 \\ 3.88 \\ 5.72 \\ 13.78 \\ 24.45} &
\makecell{3.64 \\ 4.60 \\ 6.53 \\ 10.09 \\ 16.25 \\ 25.58} &
\makecell{4.27 \\ 6.30 \\ 9.05 \\ 12.62 \\ 17.30 \\ 25.94} \\
\midrule
\makecell{BackCache \\ (12-16KB)} &
\makecell{0KB \\ 4KB \\ 8KB \\ 16KB \\ 32KB \\ 64KB} &
\makecell{3.22 \\ 3.54 \\ 3.81 \\ 3.33 \\ 2.91 \\ 2.83} &
\makecell{3.88 \\ 4.02 \\ 4.12 \\ 4.28 \\ 4.14 \\ 4.06} &
\makecell{3.54 \\ 3.79 \\ 3.91 \\ 3.77 \\ 3.60 \\ 3.63} &
\makecell{3.56 \\ 3.80 \\ 3.87 \\ 3.84 \\ 4.36 \\ 16.93} &
\makecell{3.55 \\ 3.82 \\ 3.92 \\ 3.98 \\ 8.00 \\ 24.17} &
\makecell{3.58 \\ 3.80 \\ 4.05 \\ 5.84 \\ 13.47 \\ 24.56} &
\makecell{3.70 \\ 4.82 \\ 6.58 \\ 10.30 \\ 16.23 \\ 25.83} &
\makecell{4.13 \\ 6.52 \\ 9.26 \\ 12.98 \\ 17.24 \\ 26.10} \\
\midrule
\makecell{BackCache \\ (8-16KB)} &
\makecell{0KB \\ 4KB \\ 8KB \\ 16KB \\ 32KB \\ 64KB} &
\makecell{3.23 \\ 3.58 \\ 3.82 \\ 3.47 \\ 2.93 \\ 2.83} &
\makecell{3.89 \\ 4.18 \\ 4.14 \\ 4.32 \\ 4.15 \\ 4.09} &
\makecell{3.56 \\ 3.81 \\ 3.91 \\ 3.81 \\ 3.64 \\ 4.11} &
\makecell{3.57 \\ 3.80 \\ 3.93 \\ 3.88 \\ 4.92 \\ 18.04} &
\makecell{3.60 \\ 3.82 \\ 3.99 \\ 4.01 \\ 8.67 \\ 24.26} &
\makecell{3.59 \\ 3.90 \\ 4.21 \\ 5.88 \\ 13.78 \\ 24.62} &
\makecell{3.78 \\ 4.84 \\ 6.99 \\ 10.33 \\ 16.23 \\ 25.83} &
\makecell{4.29 \\ 6.52 \\ 9.29 \\ 13.00 \\ 17.24 \\ 26.18} \\
\midrule
\makecell{BackCache \\ (4-16KB)} &
\makecell{0KB \\ 4KB \\ 8KB \\ 16KB \\ 32KB \\ 64KB} &
\makecell{3.24 \\ 3.61 \\ 3.82 \\ 3.47 \\ 2.94 \\ 2.84} &
\makecell{3.92 \\ 4.18 \\ 4.22 \\ 4.33 \\ 4.16 \\ 4.09} &
\makecell{3.56 \\ 3.84 \\ 3.94 \\ 3.83 \\ 3.65 \\ 4.21} &
\makecell{3.58 \\ 3.82 \\ 3.96 \\ 3.89 \\ 4.96 \\ 18.51} &
\makecell{3.62 \\ 3.87 \\ 4.21 \\ 4.75 \\ 8.99 \\ 24.34} &
\makecell{3.59 \\ 3.99 \\ 4.57 \\ 6.13 \\ 14.02 \\ 24.66} &
\makecell{3.96 \\ 4.93 \\ 7.06 \\ 10.82 \\ 16.33 \\ 25.84} &
\makecell{4.55 \\ 6.74 \\ 9.39 \\ 13.13 \\ 17.37 \\ 26.20} \\
\bottomrule
\end{tabular}
\begin{itemize}
\item The unit of context switch latency is µs.
\item The number of processes is 2, 4, 8, 16, 24, 32, 64, and 96.
\item The size of each process is 0KB, 4KB, 8KB, 16KB, 32KB, and 64KB.
\end{itemize}
\end{table}

\subsection{Performance Evaluation}
\begin{figure}[h!]
\centering
\includegraphics[width=\linewidth]{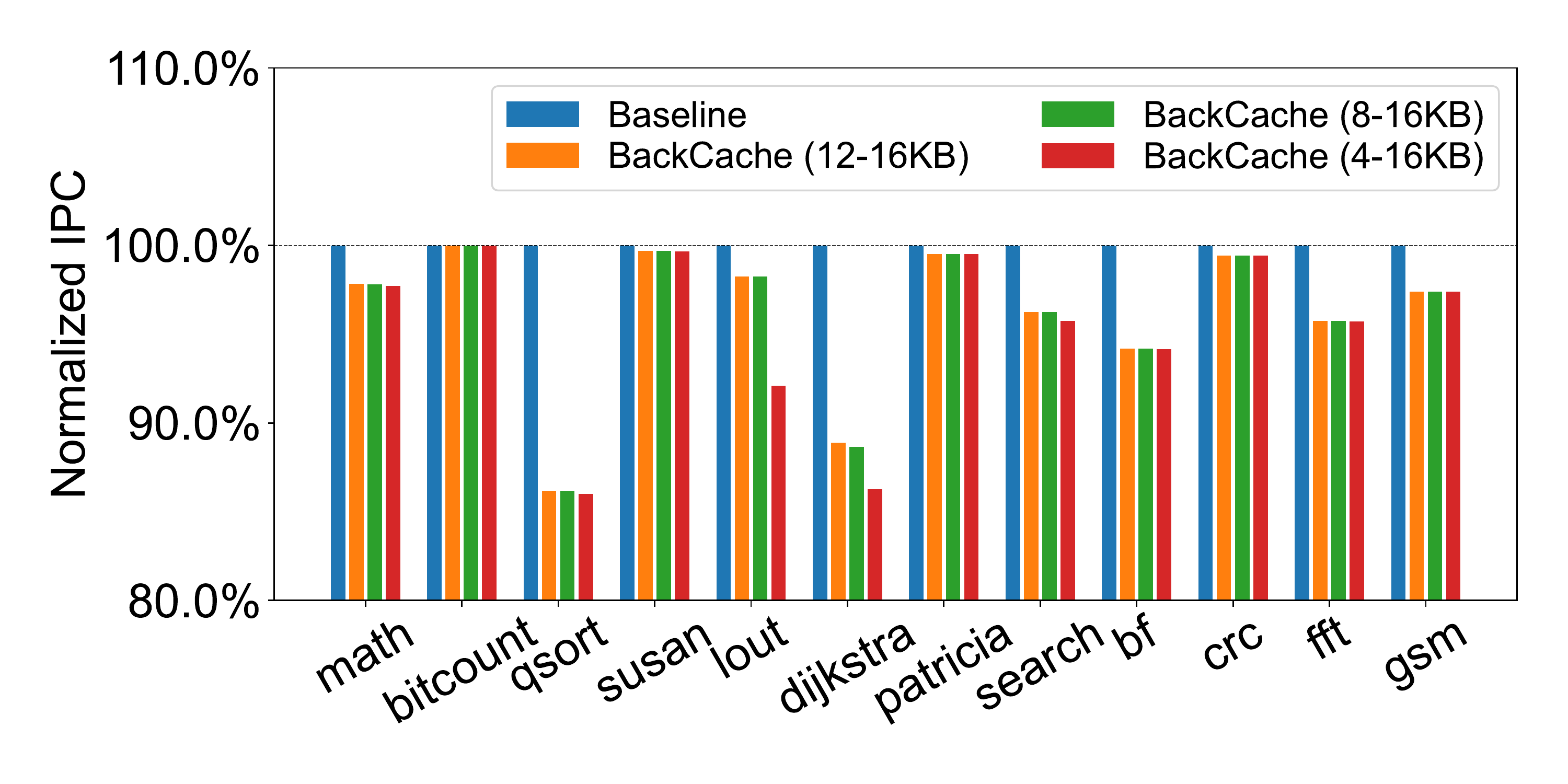}
\caption{Single-thread performance evaluation of BackCache using Mibench benchmark: normalized IPC.}
\label{fig-single-thread-ipc}
\end{figure}

\textbf{Single-Threading Performance.}
First, we evaluate the single-thread performance overhead of BackCache with
the Mibench benchmark suite in the syscall emulation mode.
We choose “large” input set for each benchmark.
Figure~\ref{fig-single-thread-ipc} and Table~\ref{tab-performance-1}
show the normalized instructions per cycle (IPC) per workload
for the baseline system and the three different BackCache systems.
The IPC of the baseline system is normalized to 1.0.
In general, the IPC of BackCache with 12-16KB dynamic backup cache decreases by an average of 3.80\%.
The IPC of BackCache with 8-16KB dynamic backup cache
and BackCache with 4-16KB dynamic backup cache decreases by an average of 3.82\% and 4.34\%.
Besides, the performance overheads of the BackCache with three different configurations
are similar for the same workload.
In the best case, all three systems have the smallest overhead on bitcount,
which is negligible for three BackCache systems.
In the worst case, all three systems have the largest overhead on qsort,
which is about 14\% for three BackCache systems.

\begin{table}[h!]
\footnotesize
\centering
\caption{The Number of IPC for Mibench Benchmark}
\label{tab-performance-1}
\begin{tabular}{ccccc}
\toprule
\textbf{Mibench} &
\textbf{Baseline} & 
\textbf{\makecell{BackCache \\ (12-16KB)}} &
\textbf{\makecell{BackCache \\ (8-16KB)}} &
\textbf{\makecell{BackCache \\ (4-16KB)}} \\
\midrule
math & 1.6385 & 1.6029 & 1.6028 & 1.6013 \\
bitcount & 3.4354 & 3.4354 & 3.4354 & 3.4354 \\
qsort & 2.2194 & 1.9127 & 1.9127 & 1.9086 \\
susan & 3.3926 & 3.3821 & 3.3821 & 3.3813 \\
lout & 0.7719 & 0.7585 & 0.7585 & 0.7109 \\
dijkstra & 2.1426 & 1.9047 & 1.8995 & 1.8483 \\
patricia & 0.8968 & 0.8925 & 0.8925 & 0.8923 \\
search & 2.0786 & 2.0009 & 2.0007 & 1.9900 \\
bf & 1.0431 & 0.9823 & 0.9823 & 0.9820 \\
crc & 1.9256 & 1.9146 & 1.9146 & 1.9146 \\
fft & 1.2754 & 1.2210 & 1.2210 & 1.2209 \\
gsm & 1.7890 & 1.7425 & 1.7425 & 1.7425 \\
\midrule
average & 1.8841 & 1.8125 & 1.8120 & 1.8023 \\
overhead & --- & +3.80\% & +3.82\% & +4.34\% \\
\bottomrule
\end{tabular}
\end{table}

\begin{table}[h!]
\footnotesize
\centering
\caption{The Number of IPC for SPEC CPU 2017 Benchmark}
\label{tab-performance-2}
\begin{tabular}{ccccc}
\toprule
\textbf{SPEC 2017} &
\textbf{Baseline} & 
\textbf{\makecell{BackCache \\ (12-16KB)}} &
\textbf{\makecell{BackCache \\ (8-16KB)}} &
\textbf{\makecell{BackCache \\ (4-16KB)}} \\
\midrule
bwaves & 0.5380 & 0.5351 & 0.5348 & 0.5343 \\
cactuBSSN & 1.2934 & 1.2543 & 1.2535 & 1.2534 \\
lbm & 0.0757 & 0.0752 & 0.0752 & 0.0752 \\
imagick & 1.4184 & 1.3109 & 1.3109 & 1.3109 \\
specrand\_f & 1.6963 & 1.6815 & 1.6815 & 1.6815 \\
perlbench & 0.7446 & 0.7209 & 0.7200 & 0.7193 \\
gcc & 0.7479 & 0.7378 & 0.7377 & 0.7362 \\
mcf & 1.6667 & 1.6024 & 1.6023 & 1.6020 \\
omnetpp & 1.9301 & 1.9146 & 1.9143 & 1.9141 \\
xalancbmk & 2.0943 & 2.0765 & 2.0763 & 2.0762 \\
deepsjeng & 0.2406 & 0.2405 & 0.2405 & 0.2405 \\
leela & 2.5073 & 2.3403 & 2.3403 & 2.3369 \\
exchange2 & 1.9179 & 1.9025 & 1.9023 & 1.9013 \\
specrand\_i & 1.6963 & 1.6815 & 1.6815 & 1.6815 \\
\midrule
average & 1.3263 & 1.2910 & 1.2908 & 1.2902 \\
overhead & --- & +2.66\% & +2.67\% & +2.71\% \\
\bottomrule
\end{tabular}
\end{table}

\begin{figure}[h!]
\centering
\includegraphics[width=\linewidth]{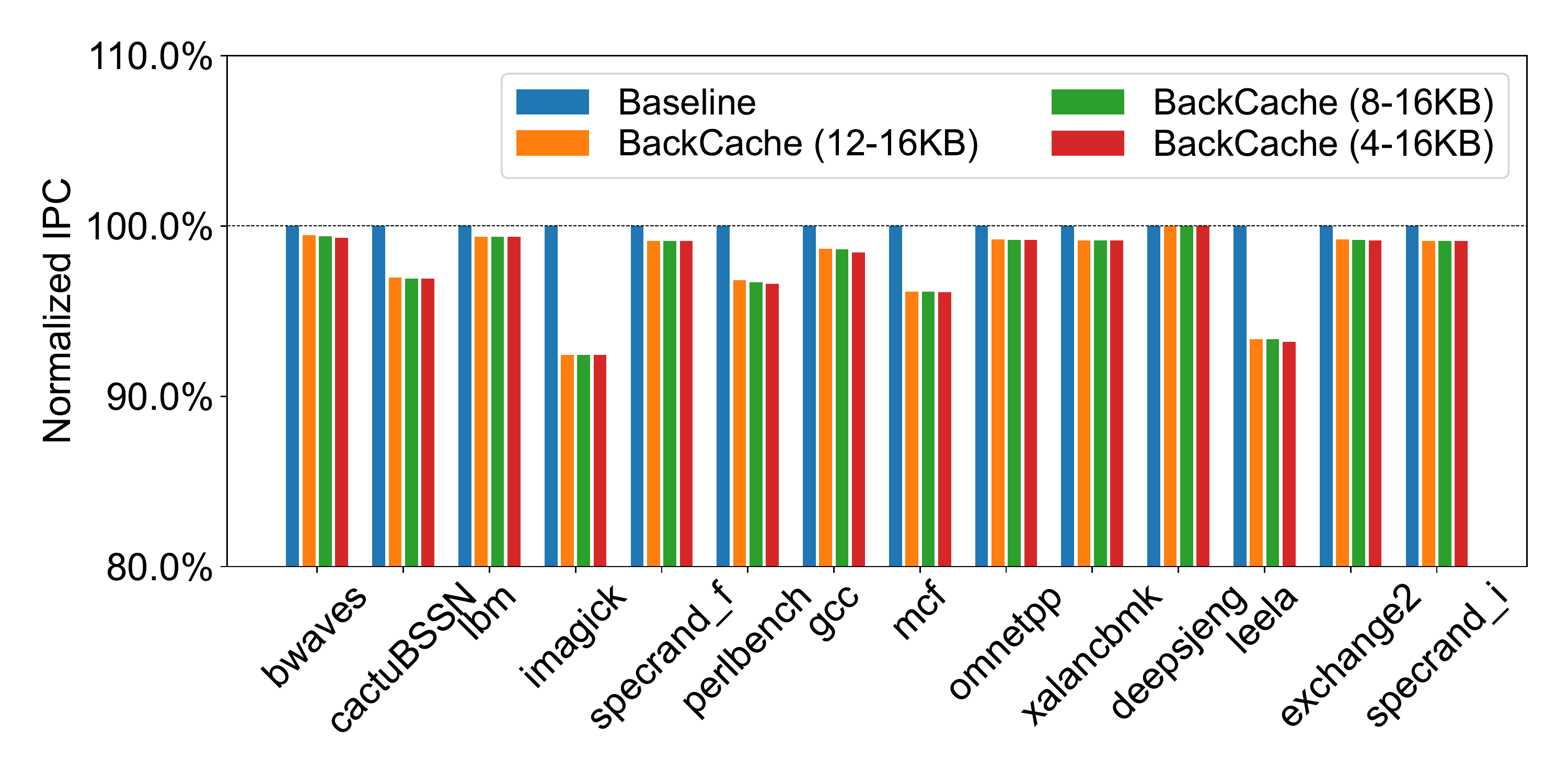}
\caption{Single-thread performance evaluation of BackCache using SPEC 2017 benchmark: normalized IPC.}
\label{fig-single-thread-ipc-2}
\end{figure}

We further evaluate the single-thread performance overhead of BackCache
using the SPEC 2017 benchmark suite with the ref input set in the system call emulation mode,
because this suite is more sensitive to L1D geometry due to its much larger footprint.
Figure~\ref{fig-single-thread-ipc-2} and Table~\ref{tab-performance-2}
show the normalized IPC per workload for
the baseline system and the three different BackCache systems.
The baseline IPC is also normalized to 1.0.
The evaluation results are similar to the Mibench benchmark suite,
where the IPC decreases by an average of 2.66\% for BackCache with 12-16KB dynamic backup cache,
by an average of 2.67\% for BackCache with 8-16KB dynamic backup cache,
and by an average of 2.71\% for BackCache with 4-16KB dynamic backup cache.
We also find that the performance overhead of
the BackCache with three different configurations for the same workload is similar.
In the best case, the IPC of deepsjeng is negligible for BackCache systems
compared to the baseline system.
In the worst case, all three systems have the highest overhead for imagick,
which is 7.58\% for three BackCache systems.

\begin{table}[h!]
\footnotesize
\centering
\caption{The Execution Time in ROI for PARSEC 3.0 Benchmark}
\label{tab-performance-3}
\begin{tabular}{ccccc}
\toprule
\textbf{PARSEC} &
\textbf{Baseline} & 
\textbf{\makecell{BackCache \\ (12-16KB)}} &
\textbf{\makecell{BackCache \\ (8-16KB)}} &
\textbf{\makecell{BackCache \\ (4-16KB)}} \\
\midrule
blackscholes & 1.238s & 1.274s & 1.275s & 1.278s \\
facesim & 8.794s & 8.951s & 8.954s & 8.977s \\
ferret & 5.465s & 5.859s & 5.865s & 5.882s \\
fluidanimate & 2.265s & 2.288s & 2.288s & 2.288s \\
freqmine & 10.100s & 10.771s & 10.779s & 10.797s \\
swaptions & 7.557s & 7.874s & 7.879s & 7.892s \\
canneal & 3.102s & 3.149s & 3.156s & 3.156s \\
streamcluster & 11.999s & 12.049s & 12.049s & 12.067s \\
\midrule
average & 6.3150s & 6.5269s & 6.5306s & 6.5421s \\
overhead & --- & +3.36\% & +3.41\% & +3.60\% \\
\bottomrule
\end{tabular}
\end{table}

\begin{figure}[h!]
\centering
\includegraphics[width=\linewidth]{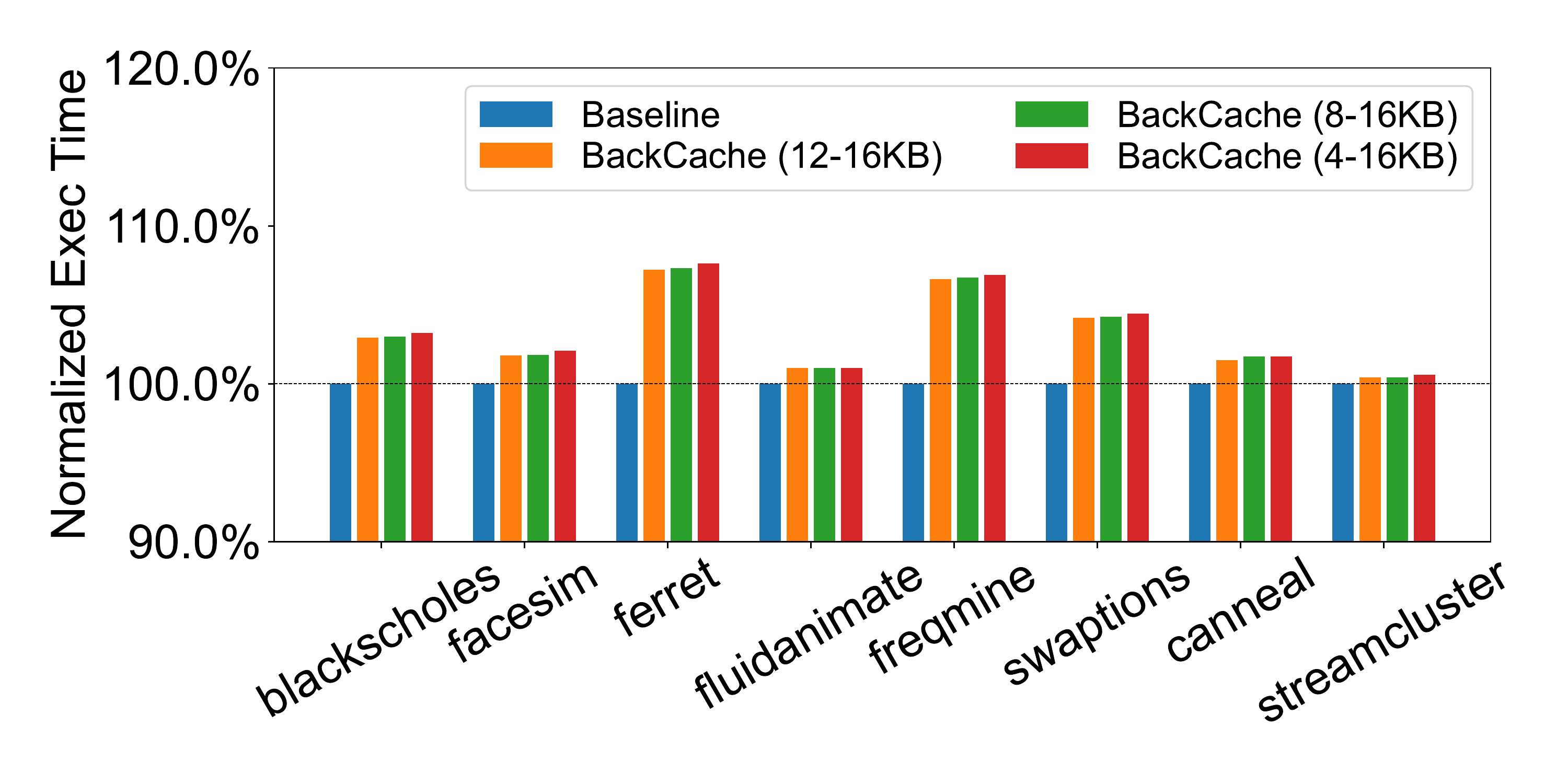}
\caption{Multicore performance evaluation of BackCache using PARSEC 3.0 benchmark: normalized execution time.}
\label{fig-multi-core-roi}
\end{figure}

\textbf{Multi-Threading Performance.}
Second, we further evaluate the multicore performance overhead of BackCache
using the PARSEC 3.0 benchmark suite in the full system mode,
and each benchmark generates four threads to accelerate the execution of the ROI on a 4-core system.
Besides, the input set for each benchmark is “simlarge”,
which is considered sufficient to measure performance without bias.
The normalized execution times in the ROI for the baseline system and the BackCache system
for each benchmark are shown in Figure~\ref{fig-multi-core-roi} and Table~\ref{tab-performance-3}.
The BackCache system with 12-16KB dynamic backup cache has an average execution time increase of 3.36\%,
and the BackCache system with 8-16KB dynamic backup cache and
4-16KB dynamic backup cache have an average execution time
increase of 3.41\% and 3.60\%, respectively.
Similar to the single-thread performance overhead,
the BackCache performance overhead is similar with
three different configurations for the same workload.
In the best case, the execution time of streamcluster decreases
by 0.42\%, 0.42\%, and 0.57\% for three BackCache systems.
In the worst case, all three systems have the highest overhead for ferret,
where the execution time increases by 7.21\%, 7.32\%, and 7.63\% on average, respectively.

\subsection{Sensitivity Analysis}
Previous experimental results indicate that the performance degradation
averages around 2.66\% for SPEC 2017 benchmarks for BackCache (12-16KB).
A key aspect of BackCache is the dynamic resizing of
backup cache capacity via the memory access count register.
Therefore, a sensitivity analysis of memory access thresholds is necessary to
strike the right balance between performance and security.

\begin{figure}[h!]
\centering
\includegraphics[width=\linewidth]{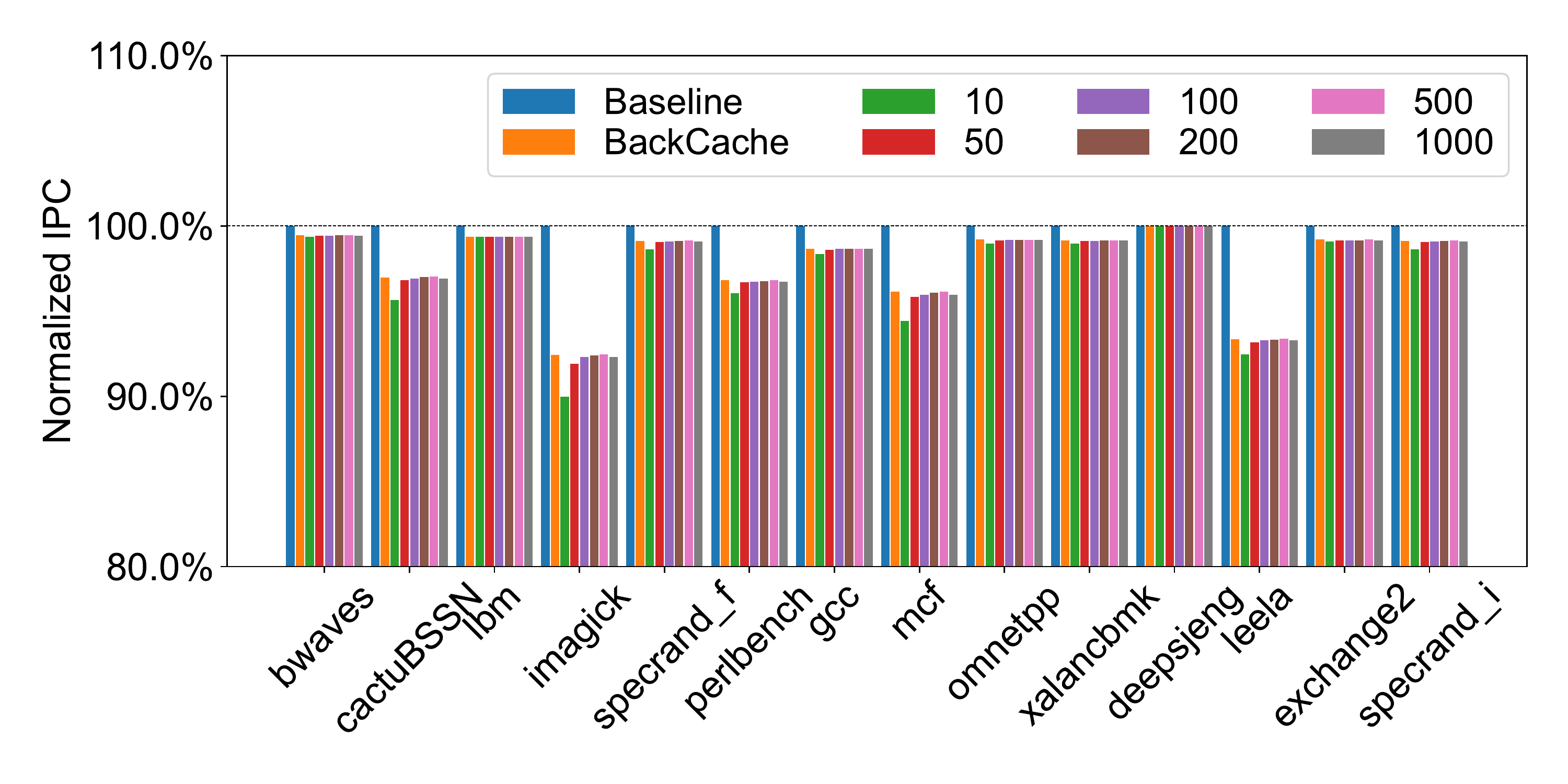}
\caption{Sensitive analysis of the memory access count threshold using SPEC 2017 benchmark: normalized IPC.}
\label{fig-sensitivity}
\end{figure}

In the sensitivity analysis experiment for memory access count,
the update thresholds for the backup cache are set to 10, 50, 100, 200, 500, and 1000,
and the performance overhead due to these thresholds is evaluated using the SPEC 2017 benchmark.
As illustrated in Figure~\ref{fig-sensitivity},
the average performance overhead reaches 3.41\% when the memory access threshold is set to 10.
Although this provides a higher level of security,
the performance overhead is 28.1\% higher than the dynamic BackCache design (12-16KB).
As the memory access count threshold increases,
the performance overhead of BackCache gradually decreases.
When the memory access threshold is set to 100, 200, and 500,
the performance overheads of BackCache are 2.71\%, 2.68\%, and 2.64\%, respectively,
which is comparable to the dynamic BackCache (12-16KB) design.
The similar performance overheads observed at thresholds of 1000
suggest that performance overheads become less sensitive to increasing thresholds beyond 500.
However, these configurations offer significantly less security
compared to the dynamic design with thresholds ranging between 192 and 256.

Overall, setting the memory access threshold to around 200 is a preferable choice
because it prevents attackers from inferring the exact backup cache capacity through memory accesses,
while keeping performance overhead relatively low.
Additionally, employing a dynamic memory access threshold
offers a security advantage by preventing attackers from detecting
when the backup cache changes,
thereby complicating their analysis of the victim's memory access patterns.

\subsection{Comparison with Related Work}
First, Düppel~\cite{zhang2013duppel} is a software solution for
refreshing cache lines from the L1 data cache.
The evaluation on real hardware shows a performance overhead
between 5\% and 7\% for most workloads in the PARSEC benchmark,
while the performance overhead of BackCache is less than 5\% for most workloads in the PARSEC benchmark.
Then, FaSe~\cite{li2022fase} is the latest hardware-based cache side-channel mitigation technique,
which selectively flushes cache lines from the L1 data cache.
Experiments conducted on a RISC-V processor show that
FaSe has an average performance overhead of 7.6\% for user programs
and incurs more than 10\% performance degradation for context switches.
Compared to this state-of-the-art cache line flushing technique,
BackCache incurs only a 2.61\% performance overhead for context switches
and results in an average 3.80\% performance degradation for the MiBench benchmark suite
(redcuing the performance overhead by 50\% compared to FaSe).

Kim et al.~\cite{kim2020revice} propose ReViCe to mitigate speculation-based timing attacks,
which usesthe idea of the victim cache to hold cache lines.
There are two major differences between ReViCe and BackCache.
First, ReViCe is designed to mitigate speculation-based timing attacks
but cannot prevent cache timing attacks,
while BackCache is designed to mitigate contention-based cache timing attacks
but cannot prevent speculation-based timing attacks.
Second, the evicted cache lines during normal execution are not kept in the victim cache in ReViCe,
while the evicted cache lines from the L1 data cache
are kept in the backup cache for further potential accesses in BackCache.

Researchers have also recently proposed some cache designs with hybrid structures.
MIRAGE~\cite{saileshwar2021mirage} extends LLC cache lines and introduces a novel storage structure
to alleviate cache timing attacks on LLC caches,
achieving security levels comparable to fully associative and randomization-based approaches.
The performance overhead imposed by MIRAGE is approximately 2\%,
with a hardware storage overhead of about 17\%-20\%.
Chameleon Cache~\cite{unterluggauer2022chameleon} is a hybrid cache design that
exploits the advantages of both the victim cache and randomization-based approaches
with a performance overhead of less than 1\%,
which has a similar idea to BackCache that uses a fully associative backup cache.
However, Chameleon Cache extends this idea only to LLC caches,
while the randomization-based approach is not feasible for L1 data caches.
ClepsydraCache~\cite{thoma2023clepsydracache} uses a novel combination of
cache decay and index randomization to mitigate the Prime+(Prune+)Probe attack.
This is an interesting randomization+ramdomization technique
that avoids the drawbacks of existing randomization techniques,
and results in an average performance overhead of only 1.38\%.

\section{Conclusion}
In this paper, we propose BackCache,
a new cache design to prevent contention-based cache timing attacks on the L1 data cache.
BackCache places the evicted cache lines into a fully associative backup cache
to achieve the goal of always having cache hits in the L1 data cache.
To improve the security of BackCache, we introduce a random used replacement policy (RURP)
and a dynamic backup cache resizing mechanism for the backup cache.
The former can prevent cache lines not reprobed by the attacker
from being evicted by leveraging the ISA extension and OS support,
while the latter based on memory access counts can prevent attacks targeting
on single cache sets and make it difficult for attackers to infer the victim cache set.
Finally, our evaluation on the gem5 simulator shows that
BackCache incurs 2.61\% performance overhead on OS kernel,
2.66\% and 3.36\% performance overhead on single-thread and multi-thread user programs.

\normalem

\bibliography{ref}
\bibliographystyle{IEEEtran}

\end{document}